\documentclass[uplatex, a4paper,12pt]{article} 
\usepackage[utf8]{inputenc}
\usepackage{amsmath}
\usepackage{color}
\usepackage{amssymb, amsfonts, amsthm, amscd}
\usepackage{bm}
\usepackage{float}
\usepackage{caption}
\usepackage{tabularx}
\usepackage{array}
\usepackage{tabularx}
\usepackage{booktabs}
\usepackage[subrefformat=parens]{subcaption}
\usepackage{algorithm}
\usepackage{algpseudocode}
\usepackage{hyperref}  
\usepackage{graphicx}  
\usepackage{bookmark}            
\hypersetup{
  setpagesize=false,
  bookmarksnumbered=true,
  bookmarksopen=true,
  colorlinks=true,
  linkcolor=blue,
  citecolor=red,
}
\newcommand{\RomNum}[1]{\uppercase\expandafter{\romannumeral #1\relax}}

\begin{document}

\thispagestyle{empty}
\vspace*{-15mm}

\begin{flushleft}
{\bf OUJ-FTC-21}\\ {\bf OCHA-PP-384}\\

\end{flushleft}

{\bf }\

\vspace{10mm}

\begin{center}
{\Large\bf
Applications of Nambu Non-equilibrium Thermodynamics to Specific Phenomena}

\baselineskip 18pt
\vspace{7mm}


So Katagiri$^{\diamond}$\footnote{So.Katagiri@gmail.com},
Yoshiki Matsuoka$^{*}$\footnote{machia1805@gmail.com}, and Akio Sugamoto$^{\dagger}$\footnote{sugamot.akio@ocha.ac.jp}

\vspace{5mm}

{\it $^{\diamond}$Major in Complex Systems Science, Graduate School of Science and Engineering,
Ibaraki University, Nakanarusawa-cho, Hitachi, 316-8511, Japan\\
\ \\
\it $^{*}$Nature and Environment, Faculty of Liberal Arts, The Open University of Japan, Chiba 261-8586, Japan\\
\ \\
\textit{$^{\dagger}$Department of Physics, Graduate School of Humanities
and Sciences, Ochanomizu University, 2-1-1 Otsuka, Bunkyo-ku, Tokyo
112-8610, Japan }

}

\end{center}

\vspace{1cm}

\setcounter{footnote}{0}
\vspace{15mm}
\begin{center}
\begin{minipage}{14cm}
\baselineskip 16pt
\noindent

\begin{abstract}
We apply Nambu non-equilibrium thermodynamics (NNET), a dynamics with multiple Hamiltonians coupled to entropy-induced dissipation, to paradigmatic far-from-equilibrium systems. Concretely, we construct NNET realizations for the Belousov-Zhabotinsky (BZ) reaction (oscillatory), the Hindmarsh-Rose neuron model (spiking), and the Lorenz and Chen systems (chaotic), and analyze their dynamical and thermodynamic signatures. Across all cases the velocity field cleanly decomposes into a non-dissipative Nambu part and a dissipative entropy-gradient part, anchored by a model-independent quasi-conserved quantity. This construction reproduces cycles, spikes, and strange-attractor behavior. These results demonstrate that NNET provides a unified, quantitatively consistent framework for oscillatory, spiking, and chaotic non-equilibrium systems, offering a systematic description beyond the Onsager-type near-equilibrium linear-response framework and complementary to nonlinear geometric formulations such as GENERIC.
\color{black}
\end{abstract}

\end{minipage}
\end{center}

\vspace{0.5cm}
\newpage
\section{Introduction}

Classical non-equilibrium thermodynamics and self-organization, ranging from de Groot–Mazur\cite{degroot1962} and the Glansdorff–Prigogine theory\cite{Prigogine} to Nicolis–Prigogine\cite{Prigogine2} and Haken's synergetics\cite{key-203}, address \color{black}dissipation\color{black}, stability, and pattern selection. The GENERIC formalism\cite{key-50,key-200,key-201,key-202} further unifies non-dissipative and dissipative \color{black} couplings within a bracket structure. In contrast, our NNET makes the \color{black}non-dissipative and dissipative\color{black}
\footnote{
  \color{black}
    In this work, ``non-dissipative'' does not imply microscopic time-reversal symmetry. 
It refers to the structural property that, after removing the entropy-gradient term, 
the remaining Nambu flow is volume-preserving (incompressible) and generates a well-posed deterministic flow (unique forward and backward trajectories) 
within the admissible state domain.
  \color{black}
}
 split explicit via Nambu brackets plus an entropy gradient. For rhythms near stable limit cycles, phase reduction originated with Winfree\cite{key-300} and Kuramoto\cite{key-301} and has since been systematized\cite{key-302,key-111}. Extensions that incorporate amplitude degrees of freedom via isostables and Koopman theory\cite{key-222,key-2222} enlarge the regime beyond weak forcing; for large oscillator ensembles, the Ott–Antonsen ansatz\cite{key-2.2} provides low-dimensional closures. Yet these approaches remain essentially local and do not furnish a thermodynamic partition of \color{black}non-dissipative\color{black} versus dissipative contributions. Spatially extended pattern formation is grounded in reaction–diffusion theory from Turing\cite{key-tur} to modern reviews\cite{key-0.1,key-0.2}. For chemical oscillations, the Belousov-Zhabotinsky (BZ) reaction is canonically treated by the Oregonator mechanism and comprehensive monographs\cite{key-0.3,key-0.4,key-0.5}. On the life-science side, spiking/bursting spans biophysically grounded models\cite{key-one}, geometric classifications\cite{key-two}, and modern syntheses\cite{key-three}. These literatures motivate a unified far-from-equilibrium decomposition, which we supply via NNET.

To address such systems, we have proposed Nambu Non-equilibrium Thermodynamics (NNET), a framework based on Nambu brackets that couples multiple Hamiltonians with entropy-induced dissipation. In this formulation, the velocity field separates cleanly into a non-dissipative Nambu component and an dissipative \color{black}   entropy-gradient component, organized by a model-independent quasi-conserved quantity (i.e., a slowly varying variable that behaves as conserved on the relevant timescale, such as
catalyst concentration in chemical oscillations or the bursting variable in neuron models; see \cite{paperI} for a formal definition).

Building on the original study ``Fluctuating Non-linear Non-equilibrium System in Terms of Nambu Thermodynamics''\cite{paper0}, which integrates \color{black}non-dissipative \color{black}   Nambu dynamics\cite{key-9} with entropy-driven dissipation, this paper further develops the framework by broadening and strengthening its applications to concrete models. As contributions of this paper, by analyzing the time evolution of representative models—the Belousov–Zhabotinsky reaction\cite{key-b,key-z} (oscillations), the Hindmarsh–Rose neuron\cite{key-51} (spiking), and the Lorenz\cite{key-lorenz} and Chen\cite{key-999} systems (chaos)—within the Nambu formalism, we demonstrate how Hamiltonian-derived non-dissipative dynamics and entropy-driven dissipation coexist and interact. This establishes that such diverse non-equilibrium behaviors can be understood at least in the paradigmatic cases studied here as manifestations of Nambu Non-equilibrium Thermodynamics (NNET), providing a unifying description beyond the Onsager-type\cite{key-3} near-equilibrium linear-response framework, while remaining complementary to nonlinear geometric approaches such as GENERIC\cite{key-200,key-201,key-202}, \color{black} in particular by accommodating regimes where entropy can locally decrease as part of the \color{black}non-dissipative \color{black} and \color{black}dissipative \color{black} interplay.

The remainder of this paper is organized as follows: Section \ref{sec2} includes a compact summary of the NNET formulation for the present 3D applications. The general case of arbitrary $N$ has already been discussed in \cite{paperII}.  Section \ref{sec3} presents practical applications and numerical analyses for the BZ reaction, the Hindmarsh–Rose\cite{key-51} model, the Lorenz system\cite{key-lorenz}, and the Chen system\cite{key-999}. Finally, in Section \ref{sec6}, we conclude with a summary and discussion of future directions.
The appendices collect supplementary materials: a minimal toy model reproducing cycle–spike transitions, a near-harmonic approximation for the BZ limit cycle, and analytical details on spike onset geometry.
\section{Approaches for Constructing the Hamiltonians and Entropy} \label{sec2}
\subsection{General evolution law of NNET}\label{sec2.1}
In this section, we briefly review the Nambu Non-equilibrium Thermodynamics (NNET) framework. NNET couples two essential flows:
 a \emph{\color{black}non-dissipative\color{black}}, volume-preserving flow generated by Nambu dynamics with multiple Hamiltonians, and a \emph{\color{black}dissipative\color{black}} flow driven by the gradient of an entropy-like potential\footnote{In Onsager's near-equilibrium theory, this $S$ coincides with the thermodynamic entropy, but in general it can differ.}($S$). This dual structure cleanly partitions the dynamics into conservative and dissipative contributions and furnishes a thermodynamic interpretation of the latter. In what follows, we first present the general evolution law (here restricted to $N=3$), then give constructive procedures for identifying the Hamiltonians and the entropy\footnote{This generalized $S$ can also be interpreted as a Lyapunov function or a dissipation potential.}($S$) for a given system, and finally illustrate the recipe on concrete models. A detailed axiomatic foundation is provided in \cite{paperI}.

We begin by recalling the basic objects of the NNET formulation and then state the general evolution law. 
In NNET, a system is described by an $N$-dimensional thermodynamic state vector 
$x=(x^1,\dots,x^N)$. The dynamics is decomposed into (i) a \emph{\color{black}non-dissipative\color{black}}, volume-preserving
flow generated by multiple Hamiltonians $H_1,\dots,H_{N-1}$ via a Nambu bracket, and (ii) an \emph{\color{black}dissipative\color{black}} flow driven by the gradient of an entropy-like potential $S$. 
In an $N$-dimensional Nambu system, the non-dissipative sector is generated by $N-1$ Hamiltonians, $H_1,\dots,H_{N-1}$.  Accordingly, the general metric-compatible NNET evolution law is\color{black}
\begin{eqnarray}\label{eq:EOM}
\dot{x}^i \;=\; -\{ H_1, H_2, \cdots, H_{N-1}, x^i \}_{\mathrm{NB}} \;+\; \sum_{j=1}^Ng^{ij}\partial_j S,
\qquad (i=1,\dots,N),
\end{eqnarray}
where the Nambu bracket $\{\cdots\}_{\mathrm{NB}}$, originally introduced by Nambu\cite{key-9} and defined as a Jacobian,
\begin{eqnarray}
\{ A_1, A_2, \cdots, A_{N} \}_{\mathrm{NB}} \equiv 
\frac{\partial(A_1,A_2,\cdots,A_N)}{\partial(x^1,x^2,\cdots,x^N)},
\end{eqnarray}
governs the non-dissipative part of the dynamics, while the gradient term represents the dissipative contribution.
Here and below, $\partial_i S$ denotes the $i$-th component of $\nabla S$.

For a general observable $O=O(x)$, the same decomposition yields
\begin{eqnarray}\label{OO}
\dot{O} \;=\; -\{ H_1, H_2, \cdots, H_{N-1}, O \}_{\mathrm{NB}} \;+\; g^{ij}\partial_i S\partial_j O.
\end{eqnarray}
Equations \eqref{eq:EOM}–\eqref{OO} define an autonomous system of $N$ ordinary differential equations,
\begin{eqnarray}\label{eq:aut}
\dot{x}^i \;=\; v^i(x), \qquad (i=1,\dots,N),
\end{eqnarray}
with velocity field $v(x)$. Since the Nambu-generated flow is divergence-free, the entropy $S$ is particularly accessible through the identity
\begin{eqnarray}\label{S}
\nabla \cdot v(x) \;=\; \partial_i(g^{ij}\partial_jS(x)).
\end{eqnarray}

It should be emphasized that the ``entropy" $S$ in NNET does not necessarily coincide with the thermodynamic entropy in the sense of Onsager's near-equilibrium theory. 
Here the word ``entropy" is used in a generalized sense, as a potential generating dissipation, rather than strictly the thermodynamic entropy. This generalized entropy inherits the role of ensuring non-negative dissipation, analogous to the second law in near-equilibrium thermodynamics.
When one of the Hamiltonians is identified with $S$ and the system is close to equilibrium, the formulation reduces to the Onsager framework. 
In general, however, $S$ should be regarded as a potential that generates dissipation and can differ from the original entropy of the nonlinear system; see \cite{paperII} for discussion. 
A deeper physical characterization of this generalized entropy via a variational principle is an important subject for future work.

Although the formulation is valid for a general $N$, in this paper we confine our analysis to the case $N = 3$, as all subsequent applications (such as the BZ reaction and the H–R model) belong to this class. In this setting, we employ two Hamiltonians, $H_1$ and $H_2$, and represent the state vector as $x = (X, Y, Z)$.

\color{black}

\subsection{Relation to metriplectic and GENERIC formulations}

It is useful to place NNET in the context of existing formulations of
non-equilibrium thermodynamics. Classical phenomenological approaches, such as
the Onsager theory\cite{key-50} and the de Groot--Mazur\cite{degroot1962} formulation, provide the standard
flux--force description of irreversible processes near equilibrium. The present
work addresses nonlinear far-from-equilibrium dynamical systems from a
geometric viewpoint, by decomposing the vector field into a non-dissipative
Nambu sector and an entropy-gradient dissipative sector.

NNET is closely related in spirit to metriplectic formulations\cite{metri}. Both approaches
combine a Hamiltonian or non-dissipative geometric contribution with a
dissipative contribution. In metriplectic systems, the non-dissipative part is
usually represented by a generalized Poisson bracket and the dissipative part
by a symmetric bracket. By contrast, in NNET the non-dissipative sector is
generated by a Nambu bracket directly on the macroscopic thermodynamic state
space. This keeps multiple Hamiltonians or quasi-conserved quantities explicit
and represents the non-dissipative part as a volume-preserving circulation,
while the dissipative part is described by an entropy-gradient flow.

NNET is also related to GENERIC\cite{key-50,key-200,key-201,key-202}. Both frameworks aim to combine a
non-dissipative geometric part with a dissipative part in a thermodynamically
consistent way. In GENERIC, the non-dissipative dynamics is formulated through
a Poisson structure, while the dissipative dynamics is formulated through a
friction matrix or, more generally, a dissipation potential, together with
degeneracy conditions\cite{ad1,ad2}. In particular, the entropy is typically required to be
a Casimir of the reversible Poisson structure. By contrast, NNET uses a Nambu
bracket with multiple Hamiltonians to generate the non-dissipative sector
directly on the macroscopic state space, while the dissipative sector is
represented by an entropy-gradient flow\footnote{\color{black}
In NNET, $S$ does not in general denote the ordinary thermodynamic
entropy $S_{\mathrm{th}}$, but rather a dissipation potential generating
the irreversible flow\cite{ad3}. The two coincide only under appropriate
conditions, such as the near-equilibrium limit corresponding to Onsager
theory. In an explicit open gas--piston model coupled to a pressure reservoir
and a heat bath, the thermodynamic entropy of the subsystem,
$S_{\mathrm{th}}$, is distinguished from the bath-relative
Massieu-type dissipation potential $S_{\mathrm{NB}} = S_{\mathrm{th}}-\frac{H_1}{T_b}$. The subsystem entropy may be nonmonotonic, whereas
$S_{\mathrm{NB}}$ is monotonic under the specified positive-semidefinite
dissipative structure\cite{ad3}.
A possible generalized formulation first uses a physical logarithmic
entropy and a state-dependent mobility to extract a generalized-gradient
component, and then decomposes the residual flow into a weighted-gradient
component and a divergence-free Nambu component.
}. \color{black}Therefore, in open far-from-equilibrium
systems, the entropy-like potential $S$ need not be a Casimir of the Nambu
bracket, and the Nambu part may contribute to the entropy rate.

\color{black}
For comparison, a purely dissipative Onsager-type gradient system,
\begin{equation}
  \dot{\boldsymbol{x}}=g\nabla S,
\end{equation}
with positive-definite $g$ satisfies
\begin{equation}
  \dot S=\nabla S^{\mathsf T}g\nabla S\geq0.
\end{equation}
Consequently, an autonomous pure-gradient system of this form cannot sustain
a nontrivial periodic orbit. This observation motivates the explicit
circulating sector used in NNET. It does not imply that GENERIC is restricted
to linear response: GENERIC may combine a Poisson structure with a nonlinear
dissipation potential\cite{ad1}, and such structures can be related to
generalized fluctuation symmetries and dynamical large-deviation
principles\cite{ad2}. The distinctive feature of NNET is instead that
multiple Hamiltonians or quasi-conserved quantities are retained explicitly
on the original macroscopic state space and generate the solenoidal flow
through a Nambu bracket. The structural distinctions are summarized in
Table~\ref{tab:framework-comparison}.

\begin{table}[t]
\color{black}
\centering
\small
\renewcommand{\arraystretch}{1.12}
\caption{\color{black}Structural comparison of representative formulations.}
\label{tab:framework-comparison}
\begin{tabularx}{\textwidth}{@{}
>{\raggedright\arraybackslash}p{0.21\textwidth}
>{\raggedright\arraybackslash}p{0.22\textwidth}
>{\raggedright\arraybackslash}p{0.22\textwidth}
>{\raggedright\arraybackslash}X@{}}
\toprule
Framework
&
Non-dissipative sector
&
Dissipative sector
&
Main distinction
\\
\midrule
Classical Onsager gradient dynamics
&
Not explicit in the pure-gradient form
&
Positive symmetric mobility acting on an entropy gradient
&
Near-equilibrium flux--force description; an autonomous pure-gradient
system cannot sustain a nontrivial periodic orbit.
\\
GENERIC\\
/metriplectic
&
Poisson or generalized Hamiltonian bracket
&
Symmetric bracket, friction matrix, or nonlinear dissipation potential
&
Multiscale thermodynamic structure with degeneracy conditions.
\\
Helmholtz--Hodge or dissipative-Nambu decomposition
&
Solenoidal or rotational vector field
&
Gradient or compressible vector field
&
General geometric decomposition; a thermodynamic interpretation requires
additional constitutive structure.
\\
NNET
&
Nambu bracket generated by multiple explicit Hamiltonians
&
Metric entropy-gradient flow
&
Retains multiple Hamiltonians or quasi-invariants on the original
macroscopic state space and provides joint $(H_1,H_2,S)$ diagnostics.
\\
\bottomrule
\end{tabularx}
\color{black}
\end{table}

\color{black}
Although Poisson-bracket-based approaches may look more general, the relation
between Nambu and Poisson descriptions is subtle. In some cases, fixing all but
one Hamiltonian in a Nambu system induces a Poisson bracket on a reduced leaf.
This, however, does not imply that the original Nambu dynamics is equivalent to
an ordinary single-Hamiltonian system on the original thermodynamic state space.
In NNET, the multiple Hamiltonians or quasi-conserved quantities are kept
explicit. These additional constraints are useful in applications because they
guide the construction of the non-dissipative, volume-preserving part of the
dynamics\footnote{\color{black}
The present formulation starts from a prescribed macroscopic evolution
equation and does not attempt a first-principles derivation of the Nambu
bracket from microscopic dynamics. In the current framework, the Nambu
structure is obtained as a local Darboux--Clebsch representation of the
solenoidal component of the effective vector field. Possible microscopic
routes through projection and coarse-graining methods remain an important
subject for future study.
}.

\color{black}
\subsection{NNET in summary (for the present 3D applications)}

We consider an $n$-dimensional state vector
$\bm{x}=(x^1,\dots,x^n)$ defined on a regular region of the state space.
Following the axiomatic formulation of NNET~\cite{paperI},
the time evolution is decomposed into a
\emph{non-dissipative} part and a \emph{dissipative} part,
\begin{equation}
\dot{x}^i
=
v_{\mathrm{non\text{-}diss}}^i(\bm{x})
+
v_{\mathrm{diss}}^i(\bm{x}),
\qquad (i=1,\dots,n).
\label{eq:Summary1}
\end{equation}

The non-dissipative part is generated by a Nambu bracket
with multiple Hamiltonians,
while the dissipative part is a gradient flow driven by
an entropy-like potential $S$ with a positive-definite
coefficient (metric) $g^{ij}$.

\vspace{1ex}

In the three-dimensional systems considered in this paper ($n=3$),
the Nambu bracket is defined as the Jacobian
\begin{equation}
\{F,G,H\}_{\mathrm{NB}}
:=
\frac{\partial(F,G,H)}{\partial(x^1,x^2,x^3)}
=
\nabla F \cdot (\nabla G \times \nabla H).
\label{eq:Summary2}
\end{equation}

The NNET evolution law used in the main text reads, in component form,
\begin{equation}
\dot{x}^i
=
- \{H_1,H_2,x^i\}_{\mathrm{NB}}
+
\sum_{j=1}^{3} g^{ij}\,\partial_j S,
\qquad (i=1,2,3)
\label{eq:Summary3}
\end{equation}
where $\partial_j S$ is a covector, and the metric or transport tensor
$g^{ij}$ maps it to the dissipative vector field. Therefore, the metric
is required for the geometrically consistent formulation of NNET. In the
explicit examples and discussions considered below, we use the Euclidean metric
$g^{ij}=\delta^{ij}$, so that the dissipative term reduces to the ordinary
gradient vector $\partial_i S$.

In the present applications, we effectively work in a Euclidean metric,
 so that the dissipative term reduces to a gradient flow; the notation with $g^{ij}$ is included only to indicate the more general form.

Here ``non-dissipative'' is used in the structural sense:
the Nambu part generates a volume-preserving
(incompressible) flow and conserves the Hamiltonians
in the absence of dissipation.
It does not necessarily imply strict time-reversal invariance
under all physical constraints.

\vspace{1ex}

The entropy rate is correspondingly decomposed as
\begin{equation}
\dot{S}
=
- \{H_1,H_2,S\}_{\mathrm{NB}}
+
\sum_{i,j=1}^{3}
(\partial_i S)\, g^{ij}\, (\partial_j S).
\label{eq:Summary4}
\end{equation}

If $g^{ij}$ is positive definite,
the second term is non-negative and represents entropy production.
The first term is not sign-definite and corresponds to
the Hamiltonian (non-dissipative) contribution to the entropy rate.
Therefore, in open or far-from-equilibrium situations,
$\dot{S}$ need not be strictly positive.
\color{black}

\subsection{Conservation laws and entropy balance}

The conservation properties of the Nambu and dissipative sectors can be made
explicit. For a Hamiltonian $H_a$, one obtains
\begin{eqnarray}
\dot H_a
=
-\{H_1,H_2,H_a\}_{\rm NB}
+
g^{ij}\partial_iH_a\,\partial_jS .
\end{eqnarray}
If $H_a$ is one of the Nambu Hamiltonians, the first term vanishes by the
antisymmetry of the Nambu bracket. Hence the Nambu sector conserves the
Hamiltonians. In the full dynamics, however, $H_a$ may change through the
entropy-gradient term unless the metric-orthogonality condition
\begin{eqnarray}
g^{ij}\partial_iH_a\,\partial_jS=0
\end{eqnarray}
is satisfied. Therefore, the Hamiltonians characterize the conservative
structure of the Nambu sector, whereas their drift in the full system measures
the coupling to dissipation or external reservoirs. \color{black}The conservation and balance identities stated above are quantitatively
verified for the BZ model in Section \ref{sec3.1}.
\color{black}
Similarly, the entropy-like potential satisfies
\begin{eqnarray}
\dot S
=
-\{H_1,H_2,S\}_{\rm NB}
+
g^{ij}\partial_iS\,\partial_jS .
\end{eqnarray}
The second term is non-negative when $g^{ij}$ is positive definite and
represents entropy production. The Nambu contribution is not sign-definite and
may represent entropy transport or entropy exchange in open systems. Therefore,
the total entropy rate need not be non-negative at every instant, although the
dissipative entropy-production term is non-negative.

In this paper, the Nambu part is referred to as non-dissipative or
volume-preserving. When the word ``reversible'' is used, it is meant only in
this structural sense: the Nambu sector does not contain the entropy-gradient
dissipation and conserves the Hamiltonians in the absence of the dissipative
term. We do not use the term to claim microscopic time-reversal invariance.
\color{black}
\subsection{Constructive procedure for Hamiltonians and entropy}\label{sec2.2}
In practice, once the general NNET evolution law has been specified, the next task is to
\emph{construct explicit forms of the Hamiltonians and the entropy} that reproduce a given set of
evolution equations. To this end, we begin with a Helmholtz-type decomposition of the velocity
field and then identify the Hamiltonian (incompressible) and entropy (gradient) parts in a
constructive manner.

The velocity field is decomposed into an incompressible part governed by Nambu dynamics and a
compressible part derived from the entropy gradient,
\begin{eqnarray}\label{eq:helmholtz}
v(x) \;=\; \nabla \times \psi \;+\; \nabla S ,
\end{eqnarray}
where, in this subsection, $\nabla$ denotes the spatial gradient (so that $\nabla\times$ and
$\nabla\cdot$ are the curl and divergence, respectively) and we work in $N=3$ dimensions.
The vector potential $\psi$ is chosen as
\begin{eqnarray}\label{eq:psi_def}
\psi \;=\; \tfrac{1}{2}\,\bigl(H_1 \overleftrightarrow{\nabla} H_2\bigr) 
\;\equiv\; \tfrac{1}{2}\,\bigl(H_1 \nabla H_2 - H_2 \nabla H_1\bigr),
\end{eqnarray}
so that $\nabla\times\psi = \nabla H_1 \times \nabla H_2$ and the incompressible component
is generated by the Nambu bracket $-\{H_1,H_2,\cdot\}_{\mathrm{NB}}$. Equation \eqref{eq:helmholtz}
thus corresponds to the standard Helmholtz decomposition into a volume-preserving
part and a gradient part. Moreover, by Darboux's theorem, the solenoidal component can
always be represented in terms of Nambu brackets with appropriately chosen Hamiltonians; see \cite{paperII} for details. Consistency with the NNET evolution law further requires the
compatibility condition $\nabla\!\cdot\!v(x)=\nabla^{2}S(x)$ (cf.\ Eq.~\eqref{S}), which we use
to determine $S$ from a given $v$.

\color{black}

\paragraph{Gauge-type non-uniqueness.}
The representation of a given velocity field in terms of $(H_1,H_2,S)$ is not unique.
If a pair $(\delta H_1,\delta S)$ satisfies
\begin{equation}
\{ \delta H_1, H_2, x^i \}_{\mathrm{NB}} + \partial_i \delta S = 0
\qquad (i=1,2,3),
\end{equation}
then the transformation
\begin{equation}
H_1 \rightarrow H_1+\delta H_1,
\qquad
S \rightarrow S+\delta S
\end{equation}
leaves the velocity field invariant.
This non-uniqueness motivates the use of explicit selection principles.

\paragraph{Selection principles.}
The gauge-type non-uniqueness described above implies that additional
\emph{selection principles} are required to identify a representation
that is physically meaningful and useful for applications.
In this work, we adopt the following criteria:

\begin{enumerate}
\item \textbf{Near- vs.\ far-from-equilibrium regime.}
When a near-equilibrium limit is relevant, we impose consistency with the
equilibrium structure in that limit (e.g., agreement with standard thermodynamic
interpretations and Onsager-type forms when applicable).

\item \textbf{Conserved or quasi-conserved quantities.}
We identify exact conserved quantities, or quantities that are approximately
conserved on the time scale of interest (pseudo-conserved variables),
and use them as natural candidates for one of the Hamiltonians.
In far-from-equilibrium systems with attractors, such quasi-invariants can often be suggested by the slow or recurrent structure of trajectories (e.g., approximate
constants of motion along the attractor), which motivates the choice of a
pseudo-conserved variable in the Nambu sector.

\item \textbf{Coupling to reservoirs and the construction of $S$.}
The dissipative potential $S$ is chosen in accordance with the type of reservoir
(thermal, chemical, mechanical, etc.) to which the system is coupled, so that
the gradient part encodes the relevant dissipation mechanism.
\end{enumerate}

Once a conserved or pseudo-conserved quantity is selected as one Hamiltonian,
the remaining Hamiltonian structure is then determined constructively through
the Helmholtz decomposition together with the Nambu representation (via Darboux's theorem).

As a simple illustration, consider a chemically open triangular reaction network
$A \leftrightarrow B \leftrightarrow C \leftrightarrow A$ coupled to chemical reservoirs.
A natural reservoir potential is the grand potential
$\Omega = F - \sum_{i\in\{A,B,C\}} \mu_i N_i$,
and conserved combinations (e.g., the total population
$N_{\mathrm{tot}} = N_A + N_B + N_C$ in a closed limit, or an approximately conserved
quantity on the relevant time scale in the open setting) provide natural candidates
for Hamiltonians\footnote{Detailed discussion of this point can be found in \cite{paperI}}.
Once such a conserved/pseudo-conserved quantity is fixed as one Hamiltonian,
the remaining Hamiltonian is then constrained by the requirement that the
Nambu (solenoidal) component reproduces the rotational part of the velocity field.

In practice, throughout this paper we adopt the simplest choice of $(H_1,H_2,S)$ that reproduces the solenoidal and gradient parts of the given vector field and is consistent with the relevant conserved or quasi-conserved quantities.
\color{black}

To simplify the construction in applications, we often adopt
\begin{eqnarray}
H_2 \;=\; Z,
\end{eqnarray}
where $Z$ is treated not as an exactly conserved quantity but as a \emph{pseudo-conserved} variable:
it remains approximately constant over the relevant timescale. For instance, in chemical reaction
systems, the concentration of a catalyst or intermediate species may oscillate periodically and serve
as such a variable. Introducing a pseudo-conserved quantity in this way frequently streamlines the
analysis of complex non-equilibrium systems.

Within this assumption, $H_1$ and $S$ can be obtained directly from the differential equations:
\begin{eqnarray}
\begin{cases}
\dot{x}^1 = \color{black}-\color{black} \partial_2 H_1 + \partial_1 S, \\
\dot{x}^2 =  \color{black}\partial_1 H_1 + \partial_2 S, \\
\dot{x}^3 = \partial_3 S.
\end{cases}
\end{eqnarray}

These lead to the following compatibility conditions between $H_1$ and $S$:
\begin{eqnarray} \label{HSCondition}
\color{black}-\color{black} \partial_2^2 H_1 \color{black}-\color{black} \partial_1^2 H_1 &=& \partial_2 \dot{x}^1 - \partial_1 \dot{x}^2, \\
\color{black}-\color{black} \partial_3 \partial_2 H_1 &=& \partial_3 \dot{x}^1 - \partial_1 \dot{x}^3, \\
\color{black} \partial_3 \partial_1 H_1 &=& \partial_3 \dot{x}^2 - \partial_2 \dot{x}^3.
\end{eqnarray}

These relations can also be obtained from the variational structure of the system. The starting point is the constitutive equation:
\begin{eqnarray}
\dot{x}^i = -\{ H_1, H_2, \dots, H_{N-1}, x^i \}_{\mathrm{NB}} + \partial_i S.
\end{eqnarray}
This can be rewritten in differential form as:
\begin{eqnarray}
dS = \delta_{ij} \left( v^i + \{ H_1, \dots, H_{N-1}, x^i \}_{\mathrm{NB}} \right) dx^j.
\end{eqnarray}

Taking the exterior derivative of both sides yields the following consistency condition:
\begin{eqnarray} \label{eq:relation}
\delta_{ij} \left( \frac{\partial v^j}{\partial x^k} + \frac{\partial}{\partial x^k} \{ H_1, \dots, H_{N-1}, x^j \}_{\mathrm{NB}} \right) dx^k \wedge dx^i = 0.
\end{eqnarray}

If the Hamiltonians are selected so as to satisfy this condition, the remaining terms in the constitutive equation can be written as an exact differential, namely, the gradient of an entropy function $S$.

We now turn to a chemical reaction system characterized by three dynamical variables, $X, Y$, and $Z$. Let $Z$ denote the concentration of a catalyst. In many such systems, the catalyst concentration exhibits periodic oscillations, which makes it natural to regard $Z$ as the second Hamiltonian:
\begin{eqnarray}
H_2 = Z.
\end{eqnarray}

Substituting this into Eq.~(\ref{eq:relation}) yields the following condition:
\begin{eqnarray}
\left(
  \color{black}-\color{black} \epsilon_{kj3} \frac{\partial^2 H_1}{\partial x^i \partial x^j}
  \color{black}+\color{black} \epsilon_{ij3} \frac{\partial^2 H_1}{\partial x^k \partial x^j}
  + \partial_k \dot{x}_i - \partial_i \dot{x}_k
\right) dx^k \wedge dx^i = 0
\end{eqnarray}
from which the relations in Eq.~(\ref{HSCondition}) immediately follow.

As we will demonstrate in subsequent sections, this condition can be explicitly satisfied in concrete models such as the BZ reaction and the Hindmarsh–Rose model.

\section{Application}\label{sec3}
In this section, we demonstrate that non-equilibrium systems far from equilibrium can be described within a unified framework of Nambu non-equilibrium thermodynamics. As illustrative cases, we analyze the BZ reaction, representing temporal oscillations as a prototypical dissipative structure, and the H–R model, which captures spike–burst behavior of the membrane potential from the perspective of NNET. We then extend the analysis to chaotic dynamical systems, examining both the Lorenz and Chen systems within the same framework.

\subsection{BZ reaction}\label{sec3.1}
As a representative periodic system, we reconstruct the Belousov–Zhabotinsky (BZ) reaction within the NNET framework and discuss the time dependence of the Hamiltonians ($H_i$) and the entropy ($S$)
The BZ reaction\cite{key-b,key-z} is a prototypical example of a non-equilibrium chemical oscillation, in which the concentrations of intermediates exhibit periodic temporal changes accompanied by striking color oscillations.
As emphasized in the pioneering works of Prigogine and coworkers, the BZ reaction has long been regarded as an archetype of dissipative structures that form far from equilibrium, providing key insights into the emergence of temporal order in chemical systems.

%
%
%
%
%
%
The simplest and most widely used reduced mechanism for the BZ reaction is the Oregonator\cite{key-0.3}. 
It idealizes the chemistry as an interplay of (i) autocatalytic production and consumption of the
activator ($X$), (ii) inhibition by bromide ($Y$), and (iii) redox cycling of the catalyst ($Z$)
driven by the organic substrate ($B$). Concretely, the elementary steps are written as
\begin{equation}
A + Y \;\xrightarrow{k_{1}}\; X + P,
\end{equation}
\begin{equation}
X + Y \;\xrightarrow{k_{2}}\; 2P,
\end{equation}
\begin{equation}
A + X \;\xrightarrow{k_{3}}\; 2X + 2Z,
\end{equation}
\begin{equation}
2X \;\xrightarrow{k_{4}}\; A + P,
\end{equation}
\begin{equation}
B + Z \;\xrightarrow{k_{5}}\; h\,Y.
\end{equation}
Here we follow the standard identification of species:
\[
A=\mathrm{BrO_3^-},\quad X=\mathrm{HBrO_2},\quad Y=\mathrm{Br^-},\quad Z=\mathrm{Ce^{4+}},\quad 
P=\mathrm{HOBr},\quad B=\mathrm{CH_2(COOH)_2},
\]
where ($k_{1},\dots,k_{5}$) are reaction rate constants and ($h$) is an adjustable
stoichiometric factor. 
In typical experimental conditions the concentrations of ($A$), ($B$), and ($P$) are maintained (buffered or in large excess) and are therefore treated as constants.

The time-evolution equations for the three concentrations $(X,Y,Z)$ are
\begin{eqnarray}
\frac{dX}{dt}&=&k_{1}AY-k_{2}XY+k_{3}AX-2k_{4}X^{2},\ \\
\frac{dY}{dt}&=&-k_{1}AY-k_{2}XY+hk_{5}BZ,\ \\
\frac{dZ}{dt}&=&2k_{3}AX-k_{5}BZ.
\end{eqnarray}
These chemical reactions are far from equilibrium and are not obtained from Onsager's variational principle.

Following the constructive procedure outlined in Section \ref{sec2.2}, we now determine the Hamiltonians $H_i$ and the entropy $S$ for the BZ reaction.
Following the methodology of Section \ref{sec2.2}, consider $Z$ as the catalyst\footnote{Here, the catalyst is interpreted as a quasi-conserved quantity.}, $H_{2}$.

Then, we obtain

\begin{equation}
d\{H_{1},Z,x^{i}\}\wedge dx^{i}=d\dot{x}^i\wedge dx^{i}.
\end{equation}

If we insert the concrete form from the development equation for velocity, $H_{1}$ must satisfy the following relation:

\begin{equation}
\begin{aligned}
\frac{\partial^{2}H_{1}}{\partial X^{2}}+\frac{\partial^{2}H_{1}}{\partial Y^{2}}&=&\color{black}-\color{black}k_{1}A\color{black}+\color{black}k_{2}(X-Y),\\
\frac{\partial^{2}H_{1}}{\partial Z\partial X}&=&hk_{5}B, \ \\
\frac{\partial^{2}H_{1}}{\partial Z\partial Y}&=&2k_{3}A. 
\end{aligned}
\end{equation}

From this,

\begin{equation}
\begin{aligned}
\frac{\partial H_{1}}{\partial X}&=&hk_{5}BZ+f(X,Y), \\
\frac{\partial H_{1}}{\partial Y}&=&2k_{3}AZ+g(X,Y), \\
\frac{\partial f(X,Y)}{\partial X}+\frac{\partial g(X,Y)}{\partial X}&=&\color{black}-\color{black}k_{1}A\color{black}+\color{black}k_{2}(X-Y).
\end{aligned}
\end{equation}
In this context, $f$ and $g$ are regarded as functions of $X$ and $Y$.
By substituting these into the above formula, 

\begin{equation}
f(X,Y)=\frac{1}{2}k_{2}X^{2}
\end{equation}
and
\begin{equation}
g(X,Y)=-k_{1}AY\color{black}-\color{black}\frac{k_{2}}{2}Y^{2}.
\end{equation}

Therefore, solving for $H_{1}$ yields

\begin{equation}
H_{1}=hk_{5}BZX-\frac{1}{2}k_{1}AY^{2}\color{black}-\color{black}\frac{k_{2}}{6}Y^{3}+2k_{3}AYZ+\frac{1}{6}k_{2}X^{3}.
\end{equation}

The time evolution of the non-Hamiltonian part is necessarily in the form of an entropy gradient.
To derive the entropy $S$, we use the decomposition of the velocity field $v$ in Eq.~(\ref{eq:helmholtz}) into an incompressible Nambu part and a gradient part, namely, $v=v^{(H)}+\nabla S$. Concretely, after determining $H_1$, we define the Nambu (Hamiltonian) contribution to the velocities by
\begin{equation}
v^{(H)} = \bigl(-\{H_1,H_2,X\}_{\mathrm{NB}},\, -\{H_1,H_2,Y\}_{\mathrm{NB}},\, -\{H_1,H_2,Z\}_{\mathrm{NB}}\bigr),
\end{equation}
and the residual by $v^{(S)}:=\dot{\bm{x}}-v^{(H)}$.
The consistency relation Eq.~(\ref{eq:relation}) ensures that this residual is an exact gradient, $v^{(S)}=\nabla S$, which is also equivalent to $\nabla\cdot v=\nabla^2 S$ in Eq.~(\ref{S}). 
Integrating $\partial_i S=v^{(S)}_i$ (up to an additive constant) yields the explicit expression\footnote{\color{black}
Here, the polynomial $S$ is an effective dissipation potential obtained
from the Helmholtz decomposition of the prescribed Oregonator vector
field and should not be identified with the physical ideal-mixture
entropy, which contains logarithmic concentration terms\cite{ad1}.
}. \color{black}This construction follows \cite{paperI,paperII}.
\begin{equation}
\begin{aligned}S= & -\frac{k_{2}}{2}XY^{2}-\frac{k_{2}}{2}X^{2}Y\\
 & -\frac{2}{3}k_{4}X^{3}+\frac{1}{2}k_{3}AX^{2}-\frac{1}{2}k_{1}AY^{2}\\
 & -\frac{k_{5}}{2}BZ^{2}+2k_{3}AXZ.
\end{aligned}
\end{equation}
\color{black}
The split  follows directly from the decomposition $v=\nabla\times\psi+\nabla S$ in Eq.~(\ref{eq:helmholtz}). 
For any observable $O$, Eq.~(\ref{OO}) reads
\begin{equation}
\dot O= -\{H_1,H_2,O\}_{\mathrm{NB}}+(\nabla S\cdot\nabla O),
\end{equation}
which we denote by $\partial_t^{(H)}O$ and $\partial_t^{(S)}O$, respectively. 
The first term is an incompressible Nambu flow that preserves phase-space volume and encodes the conservative, multi-Hamiltonian structure; the second is a compressible gradient flow generated by $S$ that accounts for dissipation and entropy production. 
This Hamiltonian--entropy split is precisely the far-from-equilibrium extension in \cite{paperI}, and it is consistent with the existence result discussed in \cite{paperII}. 
Operationally, we compute $\partial_t^{(H)}$ from $-\{H_1,H_2,\cdot\}$ and set $\partial_t^{(S)}$ by the residual $(\nabla S\cdot\nabla)$.

We divide time evolution into Hamiltonian part and entropy part: 

\begin{equation}
\frac{d}{dt}=\partial_{t}^{(H)}+\partial_{t}^{(S)}.
\end{equation}

Then, we obtain the time evolution induced by the Hamiltonians, as
follows:
\begin{eqnarray}
\partial_{t}^{(H)}X&=&-\{H_{1},H_{2},X\}_{\mathrm{NB}}=k_{1}AY+\frac{k_{2}}{2}Y^{2}-2k_{3}AZ,\ \\
\partial_{t}^{(H)}Y&=&-\{H_{1},H_{2},Y\}_{\mathrm{NB}}=\frac{k_{2}}{2}X^{2}+hk_{5}BZ,\ \\
\partial_{t}^{(H)}Z&=&-\{H_{1},H_{2},Z\}_{\mathrm{NB}}=0.
\end{eqnarray}
In the present Euclidean setting, the dissipative contribution coincides with the gradient part and can be expressed equivalently in bracket form as below.
\begin{eqnarray}
\partial_{t}^{(S)}X&=&\{S,Y,Z\}_{\mathrm{NB}}=k_{3}AX-2k_{4}X^{2}-k_{2}XY-\frac{k_{2}}{2}Y^{2}+2k_{3}AZ,\ \\
\partial_{t}^{(S)}Y&=&\{S,Z,X\}_{\mathrm{NB}}=-\frac{k_{2}}{2}X^{2}-k_{1}AY-k_{2}XY,\ \\
\partial_{t}^{(S)}Z&=&\{S,X,Y\}_{\mathrm{NB}}=2k_{3}AX-k_{5}BZ.
\end{eqnarray}
By combining these two types of time evolution, we can obtain the overall time evolution of the BZ reaction. Next, we examine the time evolution of entropy, which can be expressed as follows:

\begin{eqnarray}
\dot{S}&=&\partial_{t}^{(H)}S+\partial_{t}^{(S)}S,\ \\
\partial_{t}^{(H)}S&=&-\{H_{1},H_{2},S\}_{\mathrm{NB}},\ \\
\partial_{t}^{(S)}S&=&\{S,X,Y\}_{\mathrm{NB}}^{2}+\{S,Y,Z\}_{\mathrm{NB}}^{2}+\{S,Z,X\}_{\mathrm{NB}}^{2}.
\end{eqnarray}
The entropic part of the entropy time evolution is always positive, reflecting the essence of entropy production. In contrast, the Hamiltonian part may become negative, as it corresponds to \color{black}non-dissipative \color{black}   dynamics. Such negative contributions can arise, for example, in oscillatory reactions like the BZ reaction or in non-equilibrium steady states, where \color{black}non-dissipative \color{black}   energy or matter redistribution leads to a temporary decrease in entropy.

Now
\begin{equation}
-\{H_{1},H_{2},S\}_{\mathrm{NB}} =\color{black}\color{black}\partial_{t}^{(H)}X\partial_{t}^{(S)}X\color{black}+\color{black}\color{black}\partial_{t}^{(H)}Y\partial_{t}^{(S)}Y\color{black}+\color{black}\color{black}\partial_{t}^{(H)}Z\partial_{t}^{(S)}Z.
\end{equation}

Thus we obtain
\begin{eqnarray}
\dot{S}&=&\color{black}\partial_{t}^{(H)}X\partial_{t}^{(S)}X\color{black}+\color{black}\partial_{t}^{(H)}Y\partial_{t}^{(S)}Y\color{black}+\color{black}\partial_{t}^{(H)}Z\partial_{t}^{(S)}Z\nonumber\ \\
&+&\left(\partial_{t}^{(S)}X\right)^{2}+\left(\partial_{t}^{(S)}Y\right)^{2}+\left(\partial_{t}^{(S)}Z\right)^{2},\ \\
\dot{H}_{1}&=&\partial_{t}^{(H)}H_{1}+\partial_{t}^{(S)}H_{1},\ \\
\partial_{t}^{(H)}H_{1}&=&0.
\end{eqnarray}
Antisymmetry forces the bracket to vanish when an argument is duplicated; hence each $H_k$ is conserved along the Hamiltonian part of the dynamics.
\begin{eqnarray}
\partial_{t}^{(S)}H_{1}&=&\{H_{1},X,Y\}_{\mathrm{NB}}\{S,X,Y\}_{\mathrm{NB}}+\{H_{1},Y,Z\}_{\mathrm{NB}}\{S,Y,Z\}_{\mathrm{NB}}+\{H_{1},Z,X\}_{\mathrm{NB}}\{Z,X,S\}_{\mathrm{NB}}\nonumber\ \\
&=&\{H_{1},X,Y\}_{\mathrm{NB}}\partial_{t}^{(S)}Z+\{H_{1},Y,Z\}_{\mathrm{NB}}\partial_{t}^{(S)}X+\{H_{1},Z,X\}_{\mathrm{NB}}\partial_{t}^{(S)}Y,
\end{eqnarray}

\begin{eqnarray}
\{H_{1},X,Y\}_{\mathrm{NB}}=\color{black}k_{5}hBX+2k_{3}AY,
\end{eqnarray}

\begin{eqnarray}
\{H_{1},Y,Z\}_{\mathrm{NB}}=\frac{k_{2}X^{2}}{2}\color{black}+\color{black}k_{5}hBZ,
\end{eqnarray}

\begin{eqnarray}
\{H_{1},Z,X\}_{\mathrm{NB}}=-k_{1}AY-\frac{k_{2}Y^{2}}{2}+2k_{3}AZ,
\end{eqnarray}

\begin{eqnarray}
\dot{H}_{2}=\partial_{t}^{(H)}H_{2}+\partial_{t}^{(S)}H_{2},
\end{eqnarray}

\begin{eqnarray}
\partial_{t}^{(H)}H_{2}=0,
\end{eqnarray}

\begin{eqnarray}
\partial_{t}^{(S)}H_{2}=\partial_{t}^{(S)}Z=2k_{3}AX-k_{5}BZ.
\end{eqnarray}
The sign of the Hamiltonian time evolution can be determined as follows. The Hamiltonian part of the time evolution is given by
\begin{eqnarray}
\frac{dF}{dt}\Big|_{H} = -\{F, H_1, \dots, H_{n-1}\}_{\mathrm{NB}},
\end{eqnarray}
where the Hamiltonian flow is defined as
\begin{eqnarray}
(v_H)_i = \sum_{jk\dots l} \epsilon_{ijk\dots l} 
\frac{\partial H_1}{\partial x_j} 
\frac{\partial H_2}{\partial x_k} 
\dots 
\frac{\partial H_{n-1}}{\partial x_l}.
\end{eqnarray}
The sign of the Hamiltonian contribution to the entropy rate is determined by
\begin{eqnarray}
\mathrm{sign} \left(\frac{dS}{dt}\Big|_{H}\right) 
= \mathrm{sign} \big(-\nabla S \cdot v_H\big),
\end{eqnarray}
being positive when $-v_H$ is aligned with $\nabla S$, negative when opposed, and zero when orthogonal.
\color{black}
\paragraph{Quantitative validation of the NNET decomposition.}
We evaluated the
vector-field reconstruction, Nambu-sector orthogonality, isolated-Nambu
Hamiltonian drift, and full-system Hamiltonian balance. The parameter values
listed below are those used for the new quantitative validation.

Writing
\begin{equation}
  \boldsymbol{F}(\boldsymbol{x})
  =
  \boldsymbol{F}_{H}(\boldsymbol{x})
  +
  \boldsymbol{F}_{S}(\boldsymbol{x}),
  \qquad
  \boldsymbol{F}_{S}=\nabla S,
\end{equation}
we defined the relative vector-field reconstruction error by
\begin{equation}
  \epsilon_{\mathrm{rec}}
  =
  \left[
  \frac{
    \displaystyle
    \sum_{n=1}^{N}
    \left\|
      \boldsymbol{F}(\boldsymbol{x}_n)
      -
      \boldsymbol{F}_{H}(\boldsymbol{x}_n)
      -
      \boldsymbol{F}_{S}(\boldsymbol{x}_n)
    \right\|^2
  }{
    \displaystyle
    \sum_{n=1}^{N}
    \left\|
      \boldsymbol{F}(\boldsymbol{x}_n)
    \right\|^2
  }
  \right]^{1/2}.
\end{equation}

At each post-transient sample point $\boldsymbol{x}_n$, the normalized
orthogonality residual was defined by
\begin{equation}
  r_{a,n}^{\perp}
  =
  \frac{
    \left|
      \nabla H_a(\boldsymbol{x}_n)
      \cdot
      \boldsymbol{F}_{H}(\boldsymbol{x}_n)
    \right|
  }{
    \left\|\nabla H_a(\boldsymbol{x}_n)\right\|
    \left\|\boldsymbol{F}_{H}(\boldsymbol{x}_n)\right\|
  },
  \qquad
  a=1,2,
\end{equation}
and its root-mean-square value (RMS) was
\begin{equation}
  \epsilon_{H_a}^{\perp}
  =
  \left[
  \frac{1}{N}
  \sum_{n=1}^{N}
  \left(r_{a,n}^{\perp}\right)^2
  \right]^{1/2}.
\end{equation}

For the isolated Nambu dynamics
\begin{equation}
  \dot{\boldsymbol{x}}
  =
  \boldsymbol{F}_{H}(\boldsymbol{x}),
\end{equation}
the normalized maximum Hamiltonian drift was defined by
\begin{equation}
  D_a^{(H)}
  =
  \frac{
    \displaystyle
    \max_n\left|H_a(t_n)-H_a(0)\right|
  }{
    \displaystyle
    H_{a,\mathrm{scale}}
  },
\end{equation}
where
\begin{equation}
  H_{a,\mathrm{scale}}
  =
  \max
  \left\{
    |H_a(0)|,\,
    \max_n|H_a(t_n)|
  \right\}.
\end{equation}

For a sampled quantity $q_n$, we use
\begin{equation}
  \operatorname{RMS}(q)
  =
  \left(
  \frac{1}{N}
  \sum_{n=1}^{N}q_n^2
  \right)^{1/2}.
\end{equation}

The relative full-system Hamiltonian-balance residual was defined by
\begin{equation}
  E_a
  =
  \frac{
    \operatorname{RMS}
    \left[
      \nabla H_a\cdot\boldsymbol{F}
      -
      \nabla H_a\cdot\boldsymbol{F}_{S}
    \right]
  }{
    \max
    \left\{
      \operatorname{RMS}
      \left[
        \nabla H_a\cdot\boldsymbol{F}
      \right],
      \operatorname{RMS}
      \left[
        \nabla H_a\cdot\boldsymbol{F}_{S}
      \right]
    \right\}
  },
  \qquad
  a=1,2.
\end{equation}
This balance follows from
$\nabla H_a\cdot\boldsymbol{F}_{H}=0$ for a Nambu Hamiltonian.

\begin{table}[t]
\color{black}
\centering
\small
\caption{\color{black}Quantitative validation of the BZ NNET decomposition. The balance
residuals $E_a$ are evaluated from the chain-rule identity rather than from a
finite-difference derivative.}
\label{tab:bz-validation}
\begin{tabular}{@{}lc@{}}
\toprule
Diagnostic & Value \\
\midrule
Vector-field reconstruction, $\epsilon_{\mathrm{rec}}$
  & $1.12\times10^{-13}$ \\
$H_1$ Nambu orthogonality, $\epsilon_{H_1}^{\perp}$
  & $4.41\times10^{-18}$ \\
$H_2$ Nambu orthogonality, $\epsilon_{H_2}^{\perp}$
  & $0$ \\
$H_1$ drift in the isolated Nambu flow, $D_1^{(H)}$
  & $3.35\times10^{-15}$ \\
$H_2$ drift in the isolated Nambu flow, $D_2^{(H)}$
  & $0$ \\
$H_1$ full-system balance residual, $E_1$
  & $4.99\times10^{-14}$ \\
$H_2$ full-system balance residual, $E_2$
  & $0$ \\
\bottomrule
\end{tabular}
\end{table}

\begin{figure}[t]
\centering
\begin{subfigure}[t]{0.32\textwidth}
  \centering
  \includegraphics[width=\linewidth]{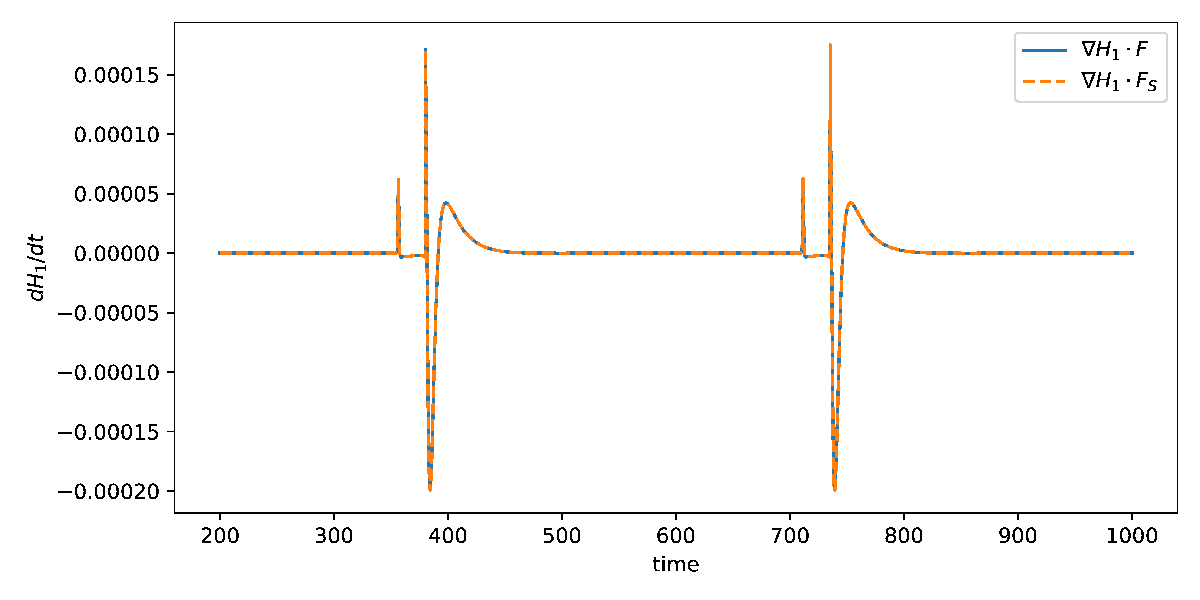}
  \caption{\color{black}Hamiltonian balance for $H_1$.}
  \label{fig:bz-H1-balance}
\end{subfigure}
\hfill
\begin{subfigure}[t]{0.32\textwidth}
  \centering
  \includegraphics[width=\linewidth]
  {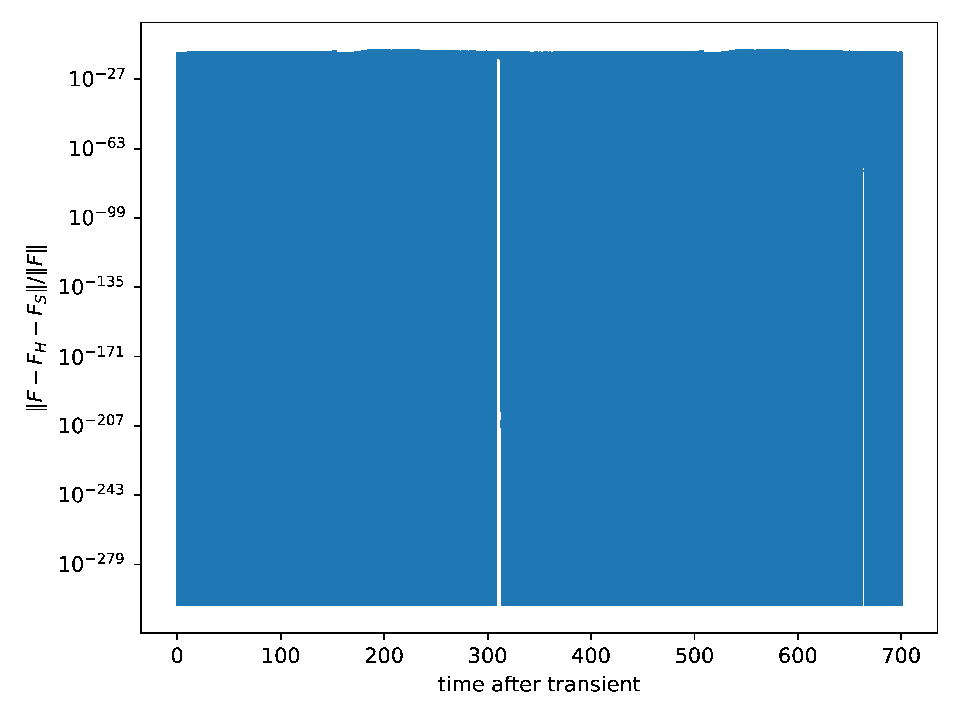}
  \caption{\color{black}Vector-field reconstruction residual.}
  \label{fig:bz-reconstruction}
\end{subfigure}
\hfill
\begin{subfigure}[t]{0.32\textwidth}
  \centering
  \includegraphics[width=\linewidth]{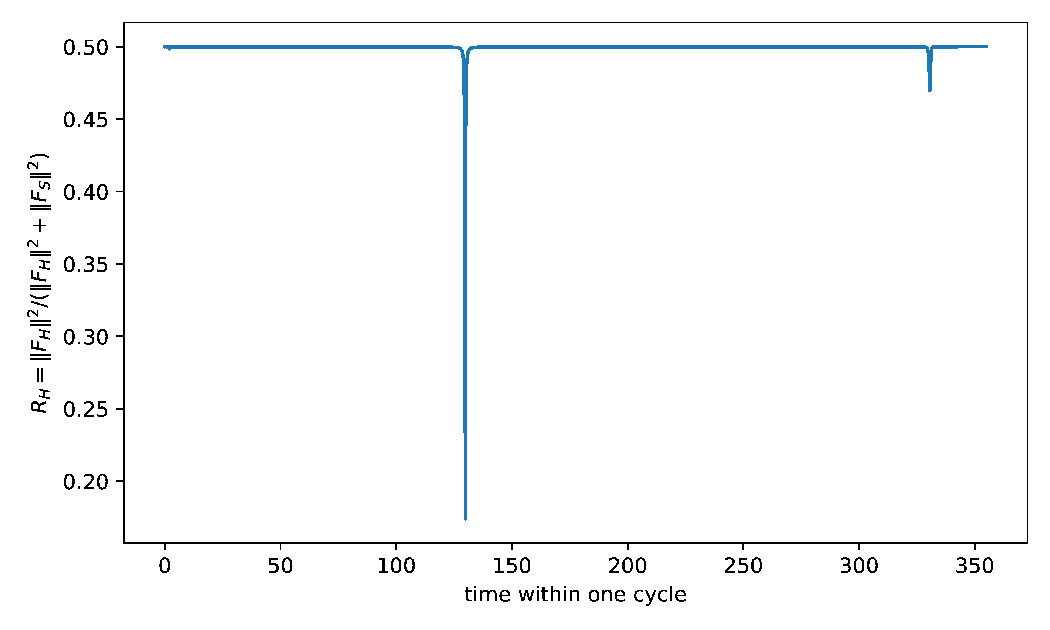}
  \caption{\color{black}Relative magnitude of the Nambu sector.}
  \label{fig:bz-Nambu-fraction}
\end{subfigure}
\caption{\color{black}Quantitative validation of the NNET decomposition for the BZ model.
(a) Comparison of $\nabla H_1\cdot\boldsymbol{F}$ and
$\nabla H_1\cdot\boldsymbol{F}_{S}$.
(b) Pointwise residual of
$\boldsymbol{F}=\boldsymbol{F}_{H}+\boldsymbol{F}_{S}$.
(c) Relative Nambu-sector magnitude
$R_H=\|\boldsymbol{F}_{H}\|^2/
(\|\boldsymbol{F}_{H}\|^2+\|\boldsymbol{F}_{S}\|^2)$ along the
post-transient trajectory.}
\label{fig:bz-quantitative-validation}
\end{figure}

Figure~\ref{fig:bz-quantitative-validation} provides a direct numerical
validation of the NNET decomposition. The two sides of the chain-rule
Hamiltonian balance for $H_1$ overlap within numerical accuracy, while the
vector-field reconstruction residual remains at the level of floating-point
error. The relative Nambu-sector magnitude further shows that the circulating
and entropy-gradient components make comparable contributions along the
oscillatory trajectory.

The results in Table~\ref{tab:bz-validation} demonstrate that $H_1$ and $H_2$
are conserved by the isolated Nambu sector within numerical accuracy, whereas
their drift in the complete dynamics is generated by the entropy-gradient
sector.

\paragraph{Sector-resolved BZ diagnostics.}
We additionally evaluated
\begin{equation}
  R_H(t)
  =
  \frac{
    \|\boldsymbol{F}_{H}\|^2
  }{
    \|\boldsymbol{F}_{H}\|^2
    +
    \|\boldsymbol{F}_{S}\|^2
  },
\end{equation}
and the directional cosine
\begin{equation}
  \cos\theta_{HS}(t)
  =
  \frac{
    \boldsymbol{F}_{H}\cdot\boldsymbol{F}_{S}
  }{
    \|\boldsymbol{F}_{H}\|
    \|\boldsymbol{F}_{S}\|
  }.
\end{equation}
The sector-resolved results are summarized in
Table~\ref{tab:bz-sector-metrics}. The Nambu and entropy-gradient vectors have
nearly equal magnitudes and are almost antiparallel. The RMS magnitudes of the
two entropy-rate contributions are individually of order unity, whereas their
sum is smaller by approximately five orders of magnitude. This is a
component-level diagnostic internal to NNET, not a performance comparison
with GENERIC or other frameworks.

\begin{table}[t]
\color{black}
\centering
\small
\caption{\color{black}Sector-resolved numerical diagnostics for the BZ trajectory.}
\label{tab:bz-sector-metrics}
\begin{tabular}{@{}lc@{}}
\toprule
Quantity & Value \\
\midrule
Mean Nambu speed fraction, $\langle R_H\rangle$
& $0.49952$ \\
Median Nambu speed fraction
& $0.499998$ \\
Mean directional cosine, $\langle\cos\theta_{HS}\rangle$
& $-0.99928$ \\
$\operatorname{RMS}(\dot S_H)$
& $1.08373118$ \\
$\operatorname{RMS}(\dot S_S)$
& $1.08373133$ \\
$\operatorname{RMS}(\dot S)$
& $4.61\times10^{-6}$ \\
$\operatorname{RMS}(\dot S)/
\max\{
\operatorname{RMS}(\dot S_H),
\operatorname{RMS}(\dot S_S)
\}$
& $4.26\times10^{-6}$ \\
\bottomrule
\end{tabular}
\end{table}
\color{black}

\begin{figure}[H]
\centering
\includegraphics[width=0.5\textwidth,height=0.3\linewidth]{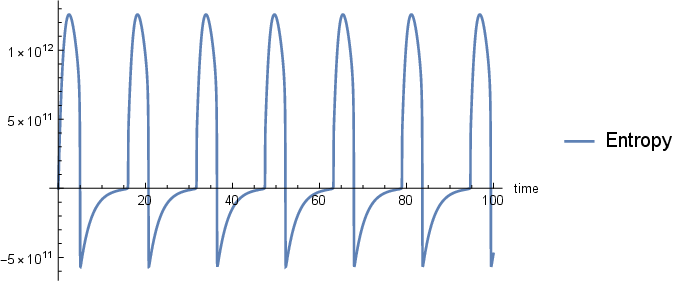}

\caption{Plot of entropy change over time. The horizontal axis represents time and the vertical axis represents entropy.\label{fig:Entropy}}
\end{figure}

\begin{figure}[H]
    \begin{tabular}{cc}
      \begin{minipage}[t]{0.45\hsize}
        \centering
        \includegraphics[width=1.0\textwidth,height=1.0\linewidth]{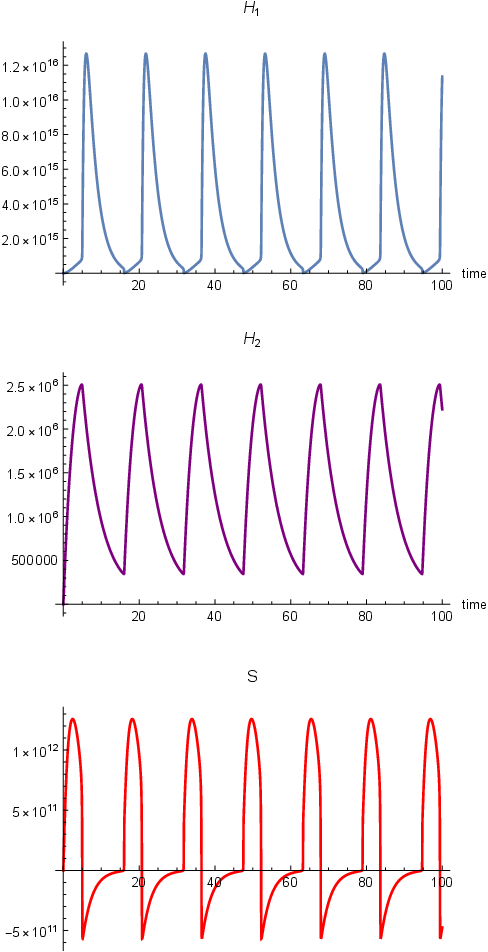}
        \caption{Plot of $H_1$, $H_2$ and $S$ as a function of time. The horizontal axis is time and the vertical axis is the values of $H_1$, $H_2$ and $S$.}
        \label{fig:H1H2S}
      \end{minipage} &
      \begin{minipage}[t]{0.45\hsize}
        \centering
        \includegraphics[width=1.0\textwidth,height=1.0\linewidth]{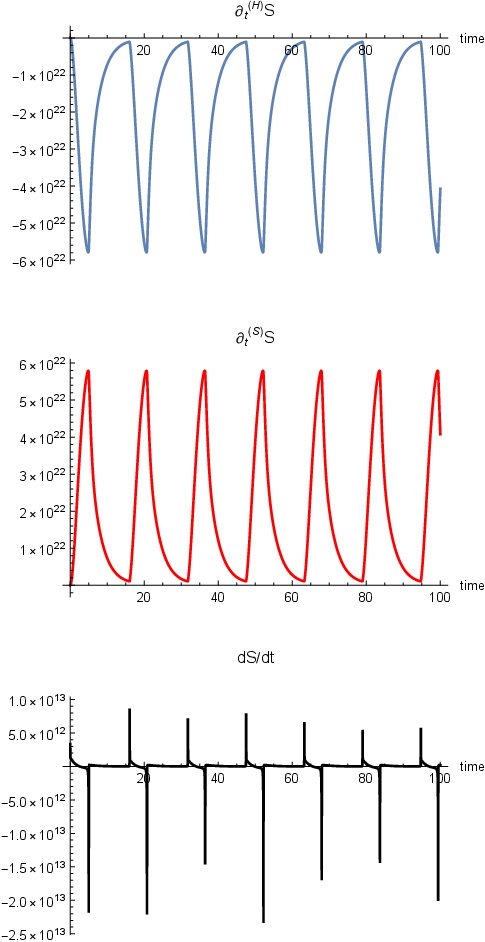}
        \caption{Plot of entropy rate as a function of time. $\partial^{(H)}_t S$ and $\partial^{(S)}_t S$ represent the contribution of the Hamiltonian in the entropy rate and the dissipation due to entropy in the time variation of entropy, respectively, and $dS/dt$ is the sum of these contributions.}
        \label{fig:entropyVelocity}
      \end{minipage}
    \end{tabular}
  \end{figure}

\begin{figure}[H]
      \begin{tabular}{cc}
        \begin{minipage}[t]{0.45\hsize}
          \centering
          \includegraphics[width=1.0\textwidth,height=1.0\linewidth]{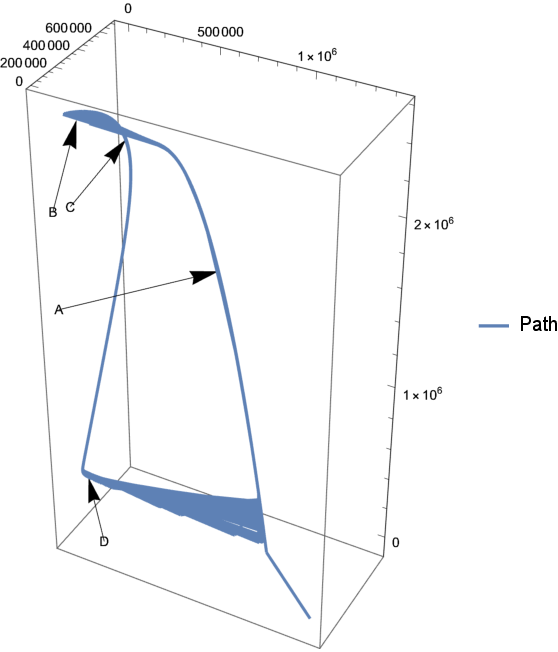}
        \end{minipage} &
        \begin{minipage}[t]{0.45\hsize}
          \centering
          \includegraphics[width=1.0\textwidth,height=1.0\linewidth]{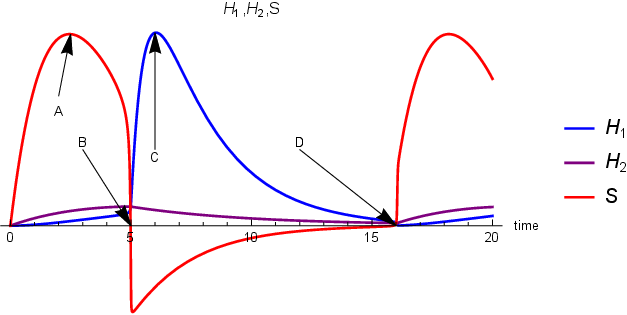}
        \end{minipage}
      \end{tabular}
      \caption{3D diagram of the limit cycle according to the time variation of the concentration X, Y and Z. The vertical axis represents the concentration of Z, the horizontal axis represents the concentration of X and the remaining axis represents the concentration of Y. \label{fig:Path}}
    
    \end{figure}

In this paper the time evolution of the BZ reaction is numerically
studied, which yields the following results depicted in Figures \ref{fig:X,Y,Z},
 \ref{fig:Entropy}, \ref{fig:H1H2S}, \ref{fig:entropyVelocity}, and \ref{fig:Path}. These figures illustrate how the BZ reaction oscillates, with the time evolution computed numerically using the Dormand–Prince method. The numerical stability was verified by tightening the relative and absolute tolerances and applying step-size control, confirming that no differences appeared in the trajectories or entropy diagnostics; the same verification was performed for the examples in the subsequent sections.

Figure \ref{fig:X,Y,Z} shows the temporal development of the
concentrations, $X$, $Y$ and $Z$, as a function of time given in
the horizontal axis. Figure \ref{fig:Entropy} gives the time development of the
entropy $S$.

Figure \ref{fig:H1H2S} gives the change of the Hamiltonian $H_{1},H_{2}$
and the entropy $S$ in time.

Figure \ref{fig:entropyVelocity} shows the temporal change of entropy
$\frac{\partial S}{\partial t}$, $\frac{\partial^{(H)}S}{\partial t}$
and $\frac{\partial^{(S)}S}{\partial t}$ in detail.

As can be seen from Figure \ref{fig:entropyVelocity}, the respective
contributions from $\frac{\partial^{(H)}S}{\partial t}$ and $\frac{\partial^{(S)}S}{\partial t}$
almost cancel each other, and entropy is found to be almost unchanged,
except for the sudden increase and decrease of the entropy, subjected
to periodic delta-function-type positive and negative kicks. The positive
and negative kicks arise alternately, where the period of positive
kick $T_{+}$ and negative kick $T_{-}$ are both $T_{+}=T_{-}\approx20$. It should also be noted that the contribution of the Hamiltonian part to the entropy transformation is always negative.

This can be seen also in the path diagram, Figure \ref{fig:Path},
where the orbits undergo an inverted kick at both ends of the elongated
circle, and only then does entropy undergo an abrupt change. It can be seen from the figure that $H_1$ and $S$ are both zero at such a turning point.
\color{black}
These results clarify why a purely autonomous Onsager-type gradient
description is insufficient to represent the closed oscillatory dynamics of
the BZ model. The point is not that nonlinear thermodynamic frameworks such
as GENERIC are unable to describe oscillations, but that a pure positive
metric-gradient flow does not contain the circulating component required for
a nontrivial periodic orbit. \color{black}The Onsager theory
is formulated near the equilibrium point of the entropy. 
There, the entropy gradient describes the time evolution of the thermodynamic variable as an affinity force, 
and as time passes, entropy increases while its rate becomes zero.
In systems far from equilibrium,  dynamics are described by entropy gradients plus Nambu dynamics.
In systems such as the BZ reaction, entropy does not necessarily increase, but rather oscillates between the Hamiltonians $H_1$ and $H_2$.

\begin{figure}[H]
  \includegraphics[width=1\textwidth,height=0.45\textwidth]{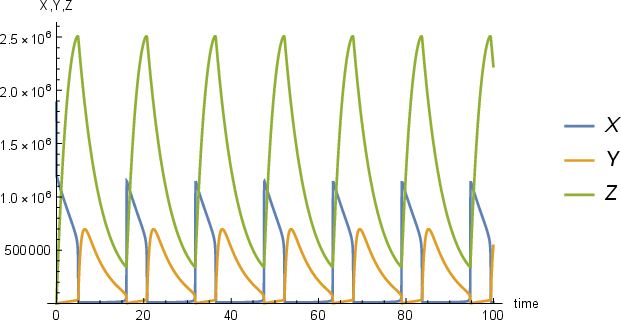}
  \begin{center}
  \caption{Time variation of concentration X, Y and Z. Horizontal axis represents time, vertical axis represents concentration.\label{fig:X,Y,Z}}
  \par\end{center}%
\end{figure}

\color{black}

\color{black}
The scaling relation used in the original manuscript,
\begin{equation}
  hk_5
  \sim
  k_1A
  \sim
  k_2
  \sim
  k_4X
  \sim
  k_5B
  \ll
  k_3A,
\end{equation}
should not be interpreted as a necessary condition for oscillation. The exact
NNET decomposition, including the expressions for $H_1$, $H_2$, and $S$, was
derived without imposing this relation. The relation is used only to motivate
the near-harmonic truncation
\begin{equation}
  H_1^{\mathrm{nh}}
  =
  2k_3AYZ,
  \qquad
  H_2=Z,
  \qquad
  S^{\mathrm{nh}}
  =
  \frac{1}{2}k_3AX^2
  +
  2k_3AXZ.
\end{equation}
\color{black}This motivates the conceptual linear oscillator-type toy model discussed in \ref{secondappendix}. It should not be regarded as a quantitatively controlled reduction of the full Oregonator limit cycle\footnote{\color{black}A rigorous local analysis of oscillation onset in the full Oregonator requires an eigenvalue analysis of the Jacobian near a stationary point. Once a periodic orbit has been established, its transverse stability must instead be examined through the Floquet multipliers obtained from the variational equations along the orbit.}.
\color{black}
Consequently, when the parameters depart from this asymptotic regime, the
exact identity
\begin{equation}
  \boldsymbol{F}
  =
  \boldsymbol{F}_{H}
  +
  \boldsymbol{F}_{S}
\end{equation}
remains valid, whereas the relative magnitudes of the retained and omitted
nonlinear terms, and hence the amplitude, period, waveform, and stability of
the motion, may change. If $Z$ ceases to be slowly varying, its physical
interpretation as a preferred quasi-conserved Hamiltonian may become less
compelling even though the algebraic decomposition remains valid.

\begin{figure}[t]
\centering
\begin{subfigure}[t]{0.48\textwidth}
  \centering
  \includegraphics[width=\linewidth]
  {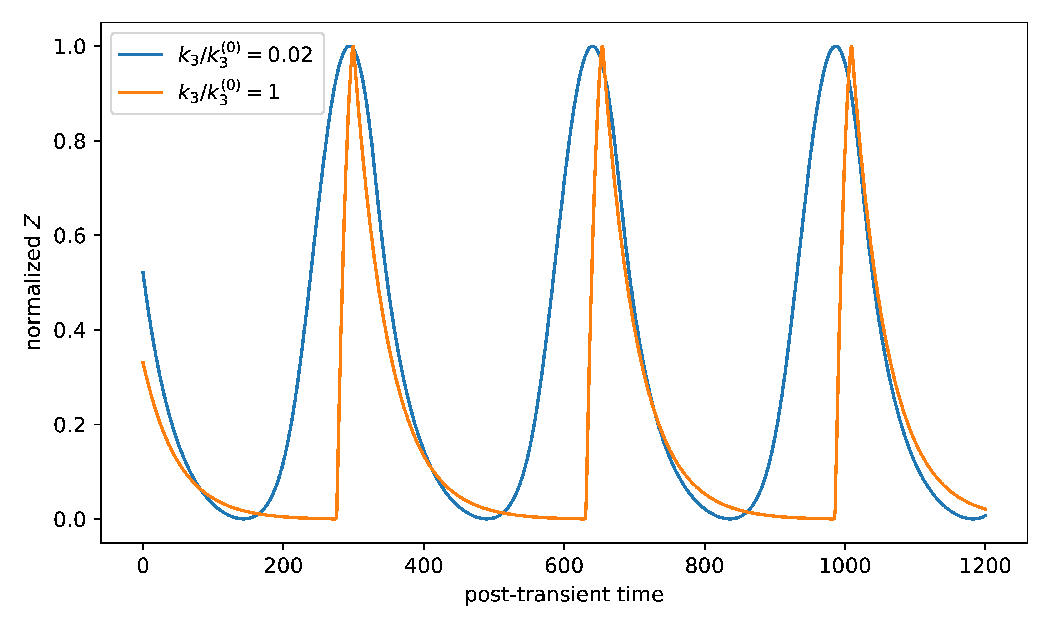}
  \caption{\color{black}Oscillation outside the assumed dominance regime.}
\end{subfigure}
\hfill
\begin{subfigure}[t]{0.48\textwidth}
  \centering
  \includegraphics[width=\linewidth]
  {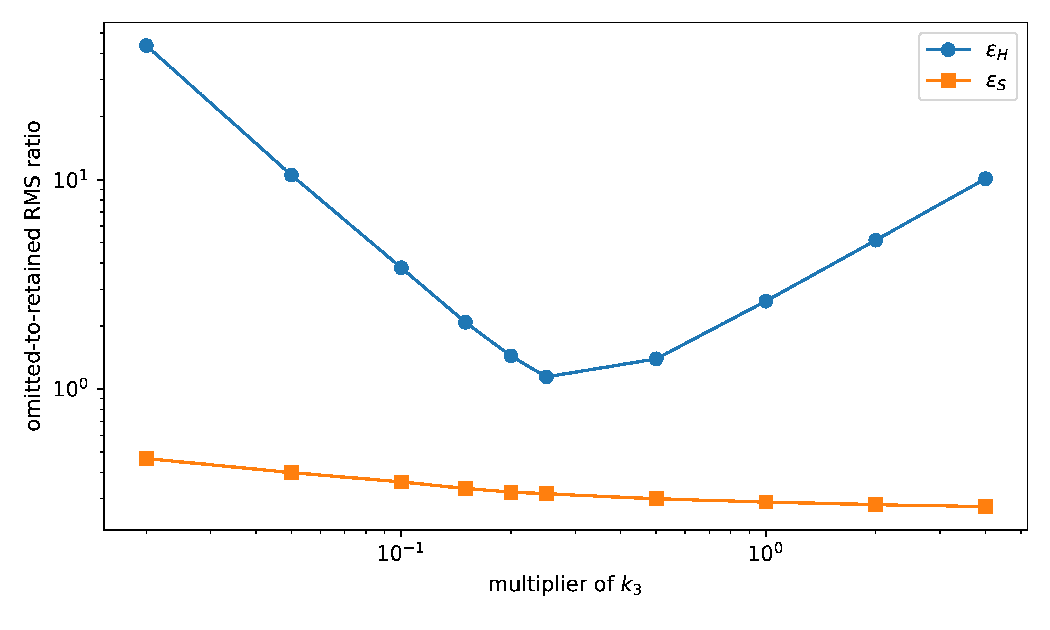}
  \caption{\color{black}Error of the near-harmonic truncation.}
\end{subfigure}

\vspace{0.5em}

\begin{subfigure}[t]{0.48\textwidth}
  \centering
  \includegraphics[width=\linewidth]
  {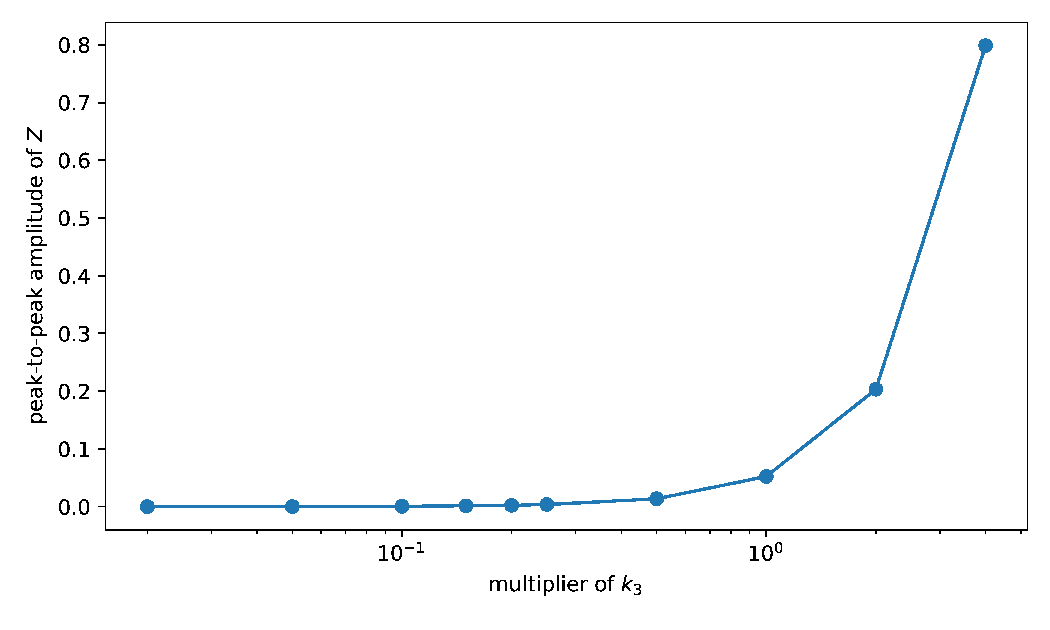}
  \caption{\color{black}Peak-to-peak oscillation amplitude.}
\end{subfigure}
\hfill
\begin{subfigure}[t]{0.48\textwidth}
  \centering
  \includegraphics[width=\linewidth]
  {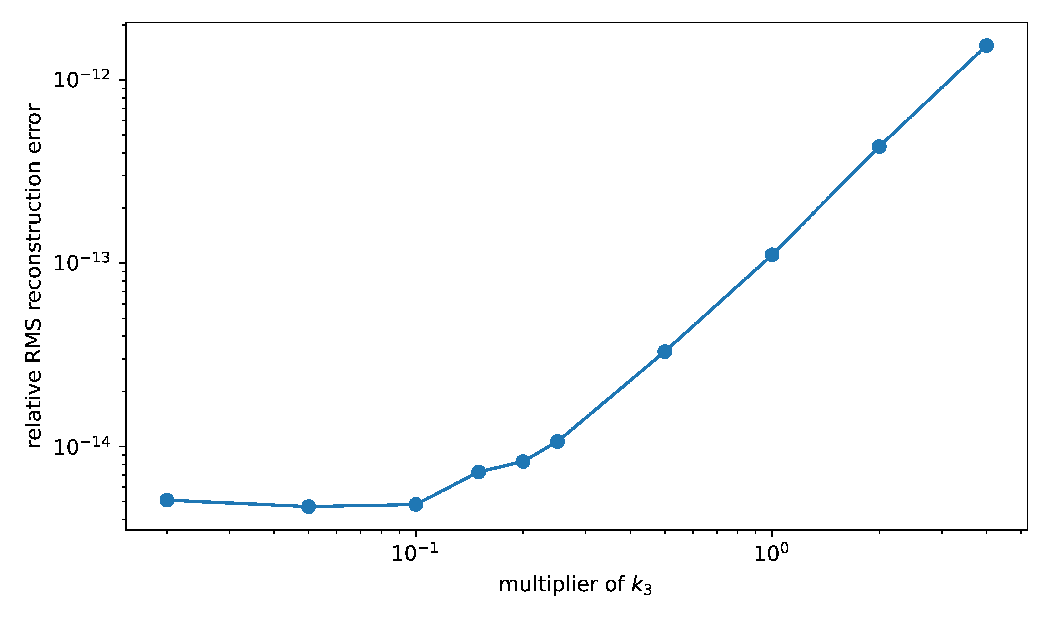}
  \caption{\color{black}Exact NNET reconstruction error.}
\end{subfigure}
\caption{\color{black}Effect of varying $k_3$ on the BZ dynamics. Numerically sustained
regular oscillations persist even when the assumed dominance of the $k_3A$
scale is violated. In this regime, the near-harmonic truncation loses
quantitative accuracy, whereas the exact NNET vector-field decomposition
remains valid within numerical precision.}
\label{fig:bz-parameter-scan}
\end{figure}

To test this distinction, we varied $k_3$ over
\begin{equation}
  0.02
  \leq
  \frac{k_3}{k_3^{(0)}}
  \leq
  4.
\end{equation}
Representative numerical values are listed in
Table~\ref{tab:bz-k3-scan}, and the results are shown in
Fig.~\ref{fig:bz-parameter-scan}.

\begin{table}[t]
\color{black}
\centering
\scriptsize
\caption{\color{black}Representative points from the $k_3$ scan. Here
$k_3=\alpha k_3^{(0)}$, $\Delta Z=Z_{\max}-Z_{\min}$,
$CV_T$ is the coefficient of variation of the peak-to-peak period,
$\epsilon_H$ and $\epsilon_S$ measure the errors of the near-harmonic
Hamiltonian and entropy-potential truncations, respectively, and
$\epsilon_{\mathrm{rec}}$ measures the exact NNET reconstruction error.}
\label{tab:bz-k3-scan}
\resizebox{\textwidth}{!}{%
\begin{tabular}{@{}rrrrrrrr@{}}
\toprule
$\alpha$
&
$k_3A$
&
$\Delta Z$
&
Mean period
&
$CV_T$
&
$\epsilon_H$
&
$\epsilon_S$
&
$\epsilon_{\mathrm{rec}}$
\\
\midrule
$0.02$
&
$6.8\times10^{-2}$
&
$2.22\times10^{-5}$
&
$346.25$
&
$7.22\times10^{-4}$
&
$43.6$
&
$0.467$
&
$5.10\times10^{-15}$
\\
$0.10$
&
$3.4\times10^{-1}$
&
$6.32\times10^{-4}$
&
$293.14$
&
$7.71\times10^{-4}$
&
$3.80$
&
$0.361$
&
$4.82\times10^{-15}$
\\
$1.00$
&
$3.4$
&
$5.22\times10^{-2}$
&
$354.92$
&
$5.25\times10^{-4}$
&
$2.63$
&
$0.289$
&
$1.11\times10^{-13}$
\\
$4.00$
&
$13.6$
&
$7.99\times10^{-1}$
&
$415.70$
&
$5.89\times10^{-4}$
&
$10.10$
&
$0.275$
&
$1.54\times10^{-12}$
\\
\bottomrule
\end{tabular}%
}
\end{table}

At $k_3/k_3^{(0)}=0.02$, the assumed dominance is violated because
\begin{equation}
  k_3A=0.068<k_1A=0.13.
\end{equation}
Nevertheless, a numerically sustained, highly regular oscillation remains over
the observation interval, with a peak-to-peak $Z$ amplitude of
$2.22\times10^{-5}$, a mean period of $346.25$, and a period coefficient of
variation of $7.22\times10^{-4}$. At the same parameter value, the
near-harmonic errors are
\begin{equation}
  \epsilon_H=43.6,
  \qquad
  \epsilon_S=0.467,
\end{equation}
whereas the exact reconstruction error is only
\begin{equation}
  \epsilon_{\mathrm{rec}}
  =
  5.10\times10^{-15}.
\end{equation}
Within the examined model and numerical range, the stated relation is
therefore not a necessary condition for the observed oscillatory behavior; it
characterizes a simplifying asymptotic regime for the near-harmonic
representation.

Oscillatory motion itself also need not require a dissipative sector. In the
pure Nambu limit, periodic motion may occur when the common level set of the
Hamiltonians forms a compact closed curve. Such conservative periodic motion
is generally neutrally stable and should be distinguished from an attracting
dissipative limit cycle, for which transverse contraction generated by the
dissipative sector is generally important.
\color{black}
    Thus, from the given Hamiltonian and entropy, we can analyze whether oscillatory phenomena occur\footnote{When the condition \(*dS = dH_1 \wedge d\Delta H_2 + d\Delta H_1\wedge dH_2 + d\Delta H_1\wedge d\Delta H_2\) is satisfied, the redefinition \(H_i \rightarrow H_i + \Delta H_i\) leads to the elimination of the entropy \(S\), leaving only the Hamiltonians to describe the system. This removal of \(S\) is crucial in achieving a quasi-conserved system of Hamiltonians. In general, the interaction between the set of Hamiltonians and entropy plays a significant role in determining the system's behavior.}.

\subsection{Hindmarsh-Rose model}
As a prototypical spiking system, we reconstruct the Hindmarsh–Rose (H–R) neuron model within NNET and discuss the time dependence of ($H_i$) and ($S$) across spike–burst cycles. The H-R model is a biological neuron model of the spike-burst behavior of the membrane potential which is a typical behavior of neurons in the brain\cite{key-51}\cite{key-52}. Understanding spikes in the H–R model enables us to describe the fundamental mechanism of neural signal transmission within the framework of dynamical systems, and moreover, to place it in a unified discussion with phenomena such as the BZ reaction as a universal behavior of non-equilibrium systems. The H-R model is a three-dimensional dynamical system, 
which is given by the following equations:
\begin{eqnarray}
\dot{x}&=&y+\phi(x)-z+I, \\
\dot{y}&=&\psi(x) - y, \\
\dot{z}&=&r(s(x - x_R)-z)
\end{eqnarray}
where
\begin{eqnarray}
\phi(x) &=& -a x^3 + b x^2, \\
\psi(x) &=& c - d x^2
\end{eqnarray}
where $x$ is the biological membrane potential and $x_R$ is the resting potential.
$y$ is called the recovery variable as it takes into account the transport of ions across the membrane through the ion channel. The recovery variable $y$ represents slower ionic processes that counteract depolarization, enabling the membrane potential to return toward rest after a spike. $z$ is the bursting variable and $I$ is the current entering the neuron from outside and is used as a control parameter.
$a,b,c,d,s$ are fixed parameters of the model and often take values such as $a=1, b=3, c=1, d=5, s=4$.
In addition, $r$ represents the timescale related to neural adaptation and assumes a very small value on the order of $10^{-3}$. The time evolution has been numerically integrated by means of the Dormand–Prince method, as in the case of the BZ reaction.

From the same discussion as for the BZ reaction, we obtain
\begin{eqnarray}
\frac{\partial^{2}H_{1}}{\partial x^{2}}+\frac{\partial^{2}H_{1}}{\partial y^{2}}&=&\color{black}-\color{black}1\color{black}-\color{black}2dx, \\
-\frac{\partial^{2}H_{1}}{\partial z\partial x}&=&0, \\
\frac{\partial^{2}H_{1}}{\partial z\partial y}&=&1+rs. 
\end{eqnarray}
From this,

\begin{eqnarray}
\frac{\partial H_{1}}{\partial x}&=&\color{black}-\color{black}f(x,y),\\
\frac{\partial H_{1}}{\partial y}&=&(1+rs)z+g(x,y), \\
\frac{\partial f(x,y)}{\partial x}+\frac{\partial g(x,y)}{\partial x}&=&2dx.
\end{eqnarray}

By substituting these into the above formula, 

\begin{eqnarray}
f(x,y)&=&\color{black}dx^{2},\ \\
g(x,y)&=&-y.
\end{eqnarray}

Therefore, $H_{1}$ is obtained as 

\begin{eqnarray}
H_{1}=-\frac{d}{3}x^{3}-\frac{1}{2}y^{2}+(1+rs)yz.
\end{eqnarray}

In the NNET, $H_1$, $H_2$, and $S$ corresponding to this model have the following forms:
\begin{eqnarray}
H_1 &=& -\frac{d}{3} x^3 - \frac{1}{2}y^2 + (1 + r s) y z, \\
H_2 &=& z, \\
S &=& - \frac{a}{4} x^4 + \frac{b}{3} x^3 + I x - \frac{y^2}{2} + c y - \frac{r z^2}{2} - r s x_R z + r s x z.
\end{eqnarray}
\color{black}
Here we set $H_2 = z$, since the bursting variable $z$ evolves on a much
slower timescale ($r \ll 1$) and can thus be regarded as a pseudo-conserved
quantity, analogous to the catalyst concentration in the BZ reaction.
Also, we divide time evolution into Hamiltonian part and entropy part as follows:

\begin{eqnarray}
\partial_{t}^{(H)}x&=&-\{H_{1},H_{2},x\}_{\mathrm{NB}}= y - (1+r s) z, \\
\partial_{t}^{(H)}y&=&-\{H_{1},H_{2},y\}_{\mathrm{NB}}= - d x^2, \\
\partial_{t}^{(H)}z&=&-\{H_{1},H_{2},z\}_{\mathrm{NB}}= 0,
\end{eqnarray}

\begin{eqnarray}
\partial_{t}^{(S)}x&=& I + b x^2 - a x^3 + r s z, \\
\partial_{t}^{(S)}y&=& c - y, \\
\partial_{t}^{(S)}z&=& r(s(x - x_R)-z).
\end{eqnarray}

Next, the time evolution of entropy is given by

\begin{eqnarray}
  \dot{S} = \partial_{t}^{(H)}S + \partial_{t}^{(S)}S, \\
\end{eqnarray}
where
\begin{eqnarray}
\partial^{(H)}_t S  &=& -d x^2 (c - y) - (I + b x^2 - a x^3 + r s z) (-y + (1 + r s) z), \\
\partial^{(S)}_t S  &=&  (c - y)^2 + r^2(s(x_R-x) + z)^2 + (I + (b - a x)x^2 + r s z)^2.
\end{eqnarray}

The time evolution of Hamiltonian is given by
\begin{eqnarray}
  \dot{H_1}&=&- r (1 + r s) y (s (x - x_R) -z) + d x^2 ( I + x^2 (b  - a x) + rsz)  \nonumber\\
  &+& (c-y) (y - (1 + r s) z), \\
  \dot{H_2} &=&r(s(x_R - x)+z).
\end{eqnarray}

\begin{figure}[H]
\includegraphics[width=1\textwidth,height=0.45\textwidth]{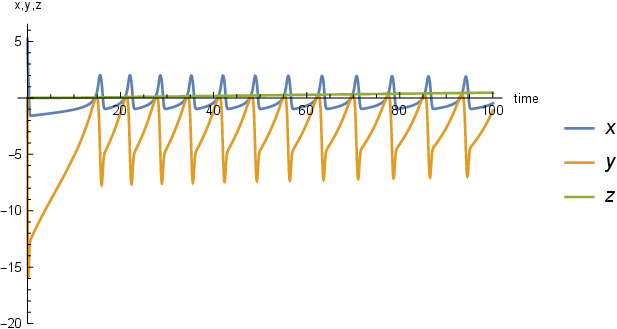}
\begin{center}
\caption{Time variation of membrane potential $x$, recovery variable $y$ and bursting variable $z$. Horizontal axis represents time, vertical axis represents $x$, $y$, $z$.\label{fig:rosexyz}}

\par\end{center}%
\end{figure}

\begin{figure}[H]
\includegraphics[width=0.5\textheight,height=0.5\textwidth]{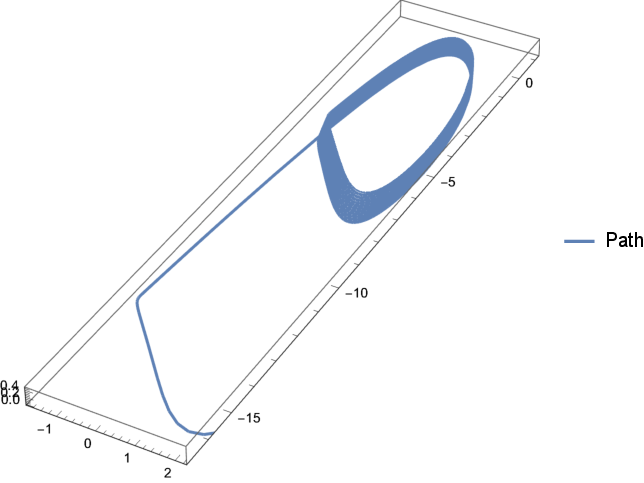}
\begin{center}
\caption{3D diagram of the limit cycle according to the time variation of $x$, $y$, and $z$. The vertical axis represents the bursting variable $z$, the horizontal axis represents the membrane potential $x$ and the remaining axis represents the recovery variable $y$. \label{fig:rosePath}}

\par\end{center}%
\end{figure}

\begin{figure}[H]
  \begin{tabular}{cc}
    \begin{minipage}[t]{0.45\hsize}
      \centering
      \includegraphics[width=1.0\textwidth,height=1.0\linewidth]{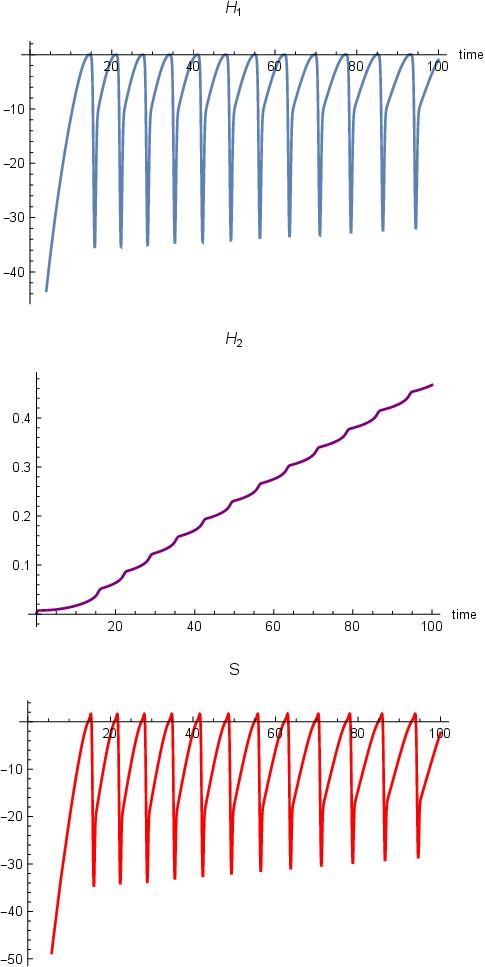}
      \caption{Plot of $H_1$, $H_2$, and $S$ as a function of time. The horizontal axis is time and the vertical axis is the values of $H_1$, $H_2$, and $S$.}
      \label{fig:roseH1H2S}
    \end{minipage} &
    \begin{minipage}[t]{0.45\hsize}
      \centering
      \includegraphics[width=1.0\textwidth,height=1.0\linewidth]{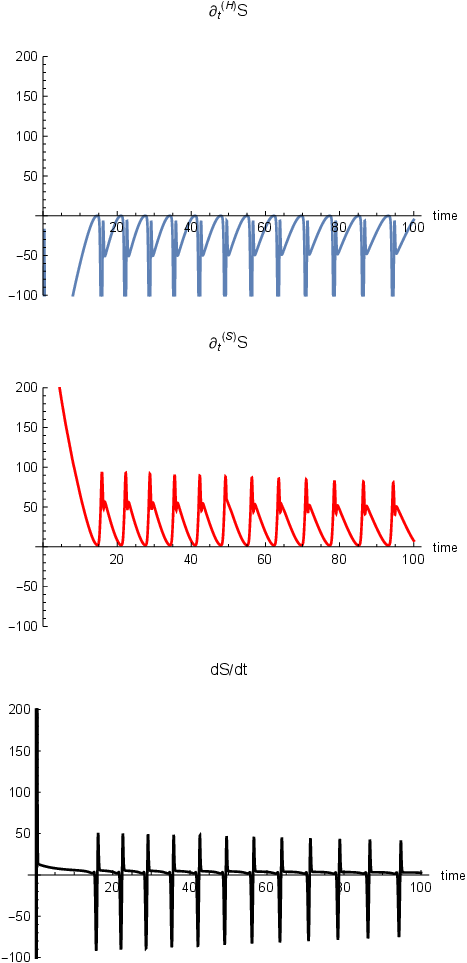}
      \caption{Plot of entropy rate as a function of time. $\partial^{(H)}_t S$ and $\partial^{(S)}_t S$ represent the contribution of the Hamiltonian in the entropy rate and the dissipation due to entropy in the time variation of entropy, respectively, and $dS/dt$ is the sum of these contributions.}
      \label{fig:roseEntropyVelocity}
    \end{minipage}
  \end{tabular}
\end{figure}

The spike-like excitatory activity of the membrane potential is shown in Figure \ref{fig:rosexyz} and Figure \ref{fig:rosePath}. The trajectory shows a limit cycle.

Figure \ref{fig:roseH1H2S}  shows the time evolution of $H_1$, $H_2$, and $S$. It can be seen that $H_1$ and $S$ take alternating peaks in a spike-like pattern, while $H_2$ increases in a staircase-like pattern.

Figure \ref{fig:roseEntropyVelocity} decomposes the time evolution of the rate of entropy into a contribution from the Hamiltonian and a contribution from entropy, with the Hamiltonian contribution of entropy having a periodic contribution to entropy.

We consider the conditions under which the time derivatives of $H_1,\ H_2,$ and $S$ are zero.

\begin{align}
\dot H_2 &= r\{\,s(x_R - x) + z\,\}=0 
\ \Longleftrightarrow\ z = s(x - x_R),
\label{Z-1} \\[2mm]
\dot S &= 0 
\ \Longleftrightarrow\ 
y = c,\quad z = s(x - x_R),\quad I + (b - a x)x^2 + rs\,z = 0,
\label{Z-2} \\[2mm]
\dot H_1 &= 0 
\ \Longleftrightarrow\ 
y = c,\quad z = s(x - x_R),\quad I + (b - a x)x^2 + rs\,z = 0.
\label{Z-3}
\end{align}

Eq.~(\ref{Z-1}) is the $z$-nullcline (balance line of the bursting variable).\ \\
Intersections of Eq.~(\ref{Z-1})–Eq.~(\ref{Z-3}) satisfy $\dot H_1=\dot H_2=\dot S=0$ and mark phase–flip points on the limit cycle (turning points or extrema of the spike velocity). Moreover, when the system is near this intersection, $H_1$ behaves as a pseudo-conserved quantity.

The staircase-like increase of $H_2$ can be understood directly from
\[
\dot H_2 = r\{z-s(x-x_R)\}.
\]
 With $r \ll 1$, so that $z$ acts as a slow variable.
  The sign of $\dot H_2$ depends on the balance between $z$ and $s(x-x_R)$; when $z > s(x-x_R)$,
   which typically occurs during spiking with $x > x_R$, we obtain $\dot H_2 > 0$.
    This mechanism leads to the staircase-like increase of $H_2$,
     representing neuronal adaptation as a slow negative feedback.

\subsection{Lorenz System}
As paradigmatic chaotic flows, we reconstruct the Lorenz and Chen systems within NNET and discuss the time dependence of ($H_i$) and ($S$) along trajectories and across regime transitions. In this subsection, we analyze the Lorenz system\cite{key-lorenz,key-t,key-j}, a representative dynamical system exhibiting chaotic behavior. The Lorenz system is widely known as a simplified model of convection and fluid dynamics.

The Lorenz system can be described as a toy model consisting of the following set of three differential equations:
\begin{eqnarray}
\dot{x} &=& \sigma(y - x), \\
\dot{y} &=& x(\rho-z) - y, \\
\dot{z} &=& xy - \beta z,
\end{eqnarray}
where we set the parameters as $\sigma = 10$, $\beta = \frac{8}3$ with the time evolution computed numerically using the Dormand–Prince method.

\begin{figure}[H]
\centering
\includegraphics[width=1.0\textwidth,height=0.45\textwidth]{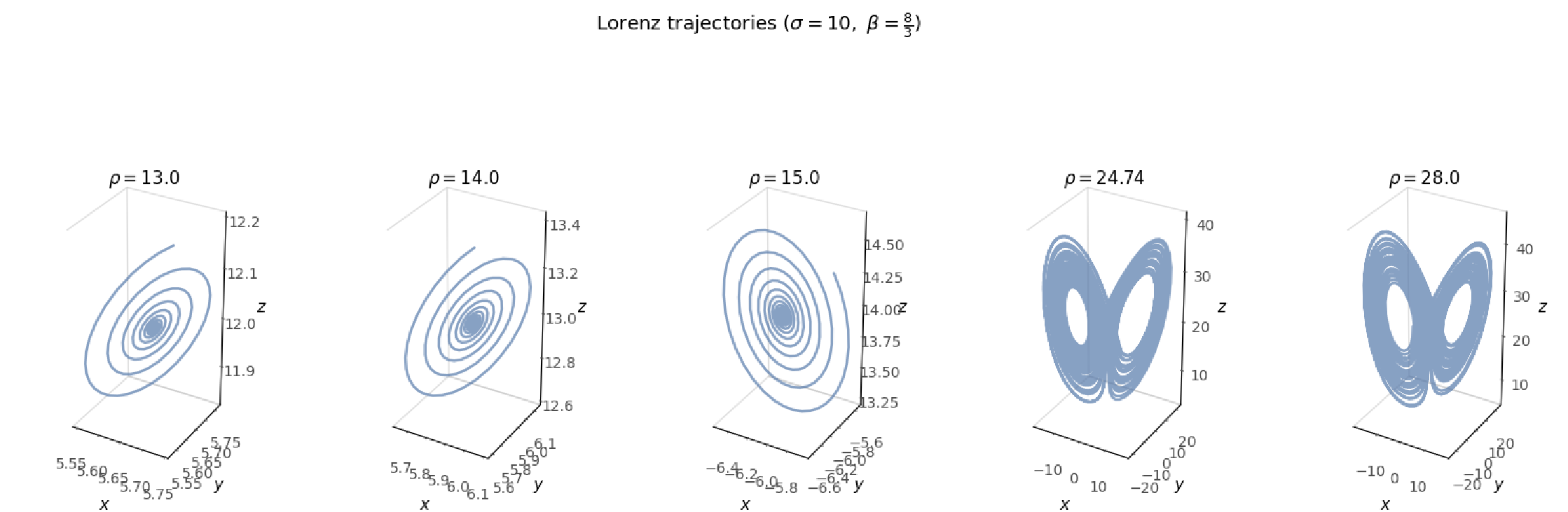}
\caption{Trajectory of the Lorenz system with initial condition $x(0) = 1.0$, $y(0) = 1.0$, $z(0) = 1.0$, and $dt_{init} = 0.005$. Displayed side by side for increasing values of $\rho$ ($\rho=13.0,\ 14.0,\ 15.0,\ 24.74,\ 28.0$): \color{black}the cases $\rho=13$, $14$, and $15$ show damped oscillations toward stable
nonzero equilibria, whereas $\rho=24.74$ and $28$ exhibit chaotic dynamics
for the stated initial condition.\color{black}
}
\label{a}
\end{figure}
Figure \ref{a} shows the typical attractor of the Lorenz system in three-dimensional phase space. The two circular lobes correspond to trajectories revolving around the two nonzero equilibria. Chaos arises when the switching between these lobes becomes aperiodic, which can be quantified in the NNET framework: on the Poincaré section, the sequence $\{H_1(t_k),S(t_k)\}$ evolves from clustered to band-like and eventually to diffuse distributions.

Within the framework of NNET, we introduce the following expressions for the Hamiltonians $H_1$, $H_2$, and the entropy $S$\footnote{Axenides and Floratos (A\&F)\cite{AF} study dissipative dynamics by adopting a Helmholtz decomposition of the vector field (rotational part = non-dissipative, gradient part = dissipative), which is formally compatible with the type of splitting used in our discussion. Our NNET framework, however, is formulated as a non-equilibrium thermodynamics: the \color{black}dissipative \color{black}   contribution is required to take the thermodynamic form $g^{ij}\nabla_j S$ with a non-negative dissipation metric, so that the structure of $\dot{S}$ explicitly separates entropy exchange from non-negative entropy production. In contrast, the gradient term in A\&F is introduced primarily to represent phase-space volume contraction and is not tied to such thermodynamic constraints. Finally, the \color{black}non-dissipative/dissipative \color{black}   separation itself is generally non-unique; in NNET this is reflected, for example, in the reparameterization freedom of $H_1,H_2$ (e.g., an $SL(2,\mathbb{R})$-type freedom). More concretely, this redundancy allows one to reshuffle contributions between the Nambu and gradient parts by a compensating transformation of the form $0=\{x^i,H_1,Q\}_{\mathrm{NB}}+\partial_i(\delta S)$ (for a suitable pair $(Q,\delta S)$), which leaves $\dot{x}^i$ unchanged while modifying the split. Accordingly, our representation $(H_1,H_2,S)$ is written so as to retain this redundancy.}:
\begin{eqnarray}
H_1 &=& \frac{1}{2}(-\rho x^2+\sigma y^2+\sigma z^2), \\
H_2 &=& z-\frac{1}{2\sigma}x^2, \\
S &=& -\frac{1}{2}(\sigma x^2 + y^2 + \beta z^2).
\end{eqnarray}

\begin{figure}[H]
\centering
\includegraphics[width=0.65\textwidth,height=0.45\textwidth]{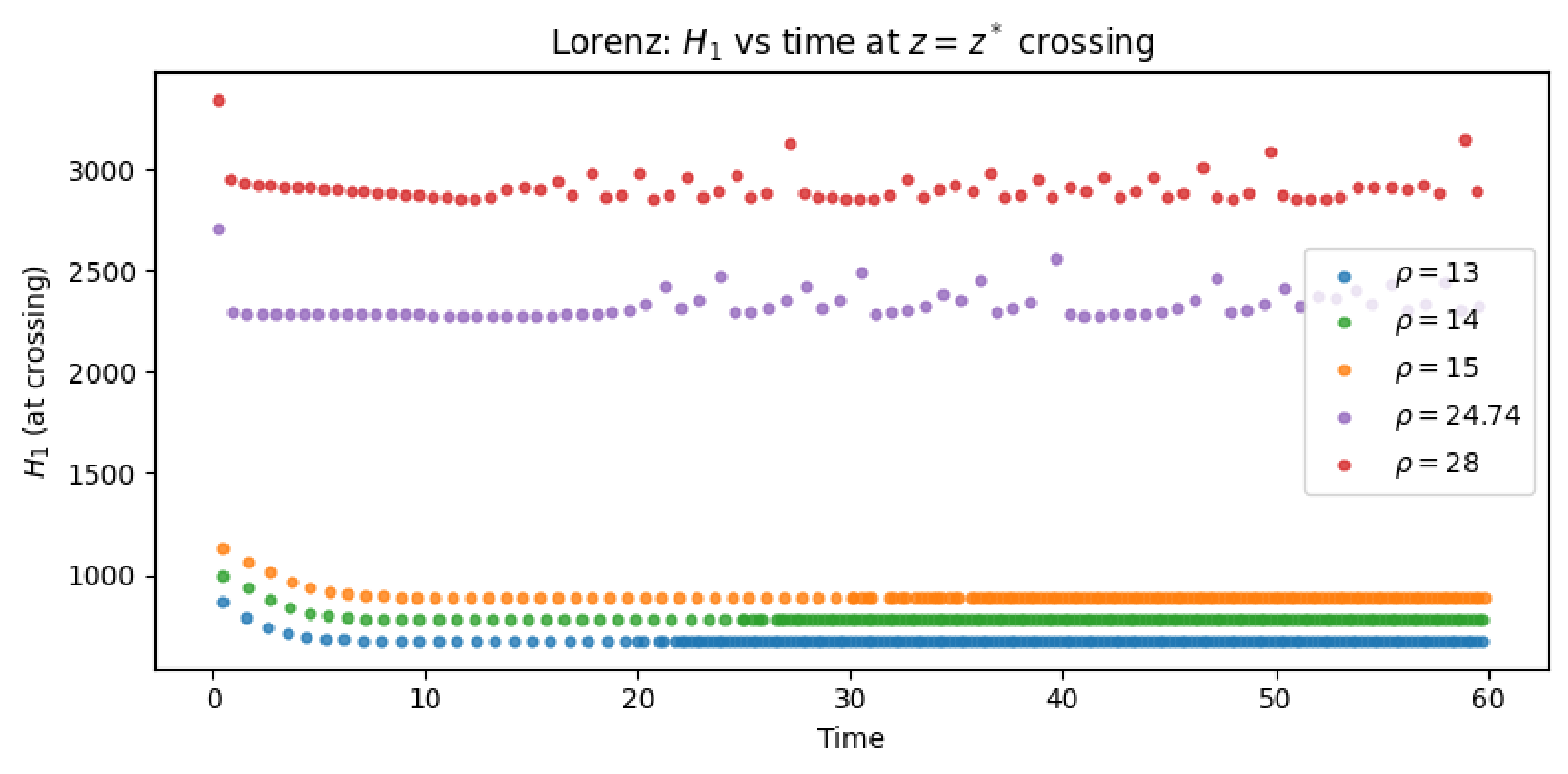}
\caption{Time evolution of the Hamiltonian $H_1$ for the Lorenz system over the time interval $0 \le t \le 60$ and $dt_{init} = 0.005$ with initial condition $x(0) = 1.0$, $y(0) = 1.0$, and $z(0) = 1.0$.}
\label{b}
\end{figure}

\begin{figure}[H]
\centering
\includegraphics[width=0.65\textwidth,height=0.45\textwidth]{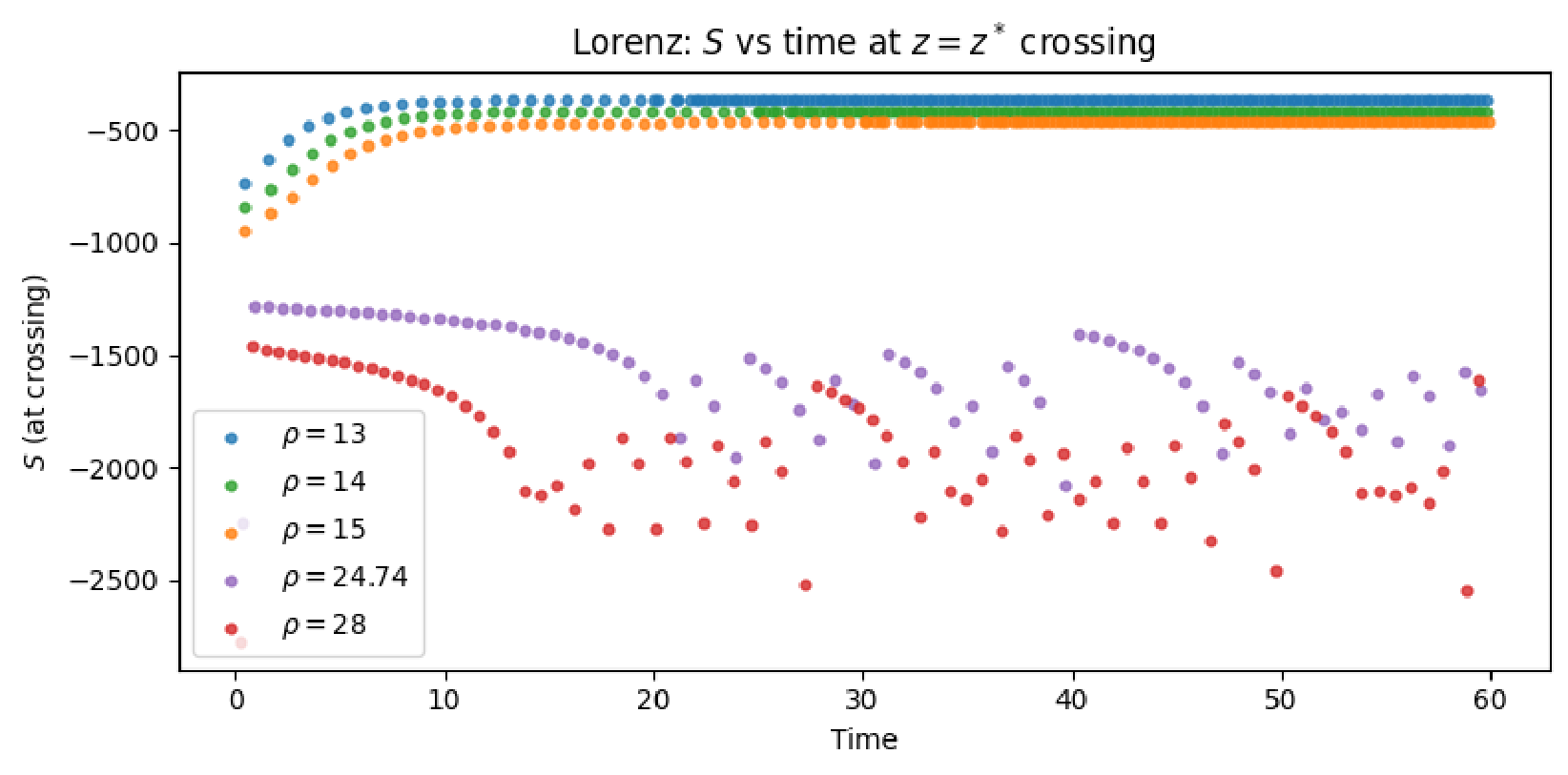}
\caption{Time evolution of the Hamiltonian $S$ for the Lorenz system over the time interval $0 \le t \le 60$ and $dt_{init} = 0.005$ with initial condition $x(0) = 1.0$, $y(0) = 1.0$, and  $z(0) = 1.0$. 
}
\label{c}
\end{figure}
We next apply the NNET section analysis to the Lorenz system. Figures~\ref{b} and \ref{c} display the time evolution of $H_1$ and $S$ over the interval $0 \le t \le 60$ with initial condition $x(0)=y(0)=z(0)=1.0$ and $dt_{init}=0.005$. 
To extract the underlying structure, we sample the trajectory at times $t_k$ when it intersects the section $z=z^\ast:=\rho-1$, corresponding to the $z$-coordinate of the nonzero equilibria, and record the values $H_1(t_k)$ and $S(t_k)$.

The sampled sequences reveal a clear progression: in the stationary regime the points remain tightly clustered, in the periodic regime they form band-like structures reflecting alternation between the two lobes, and in the chaotic regime they spread broadly across the section. 
Thus, the two circular lobes in the phase space are directly connected to these rearrangements in the $(H_1,S)$ plane.

Moreover, near bifurcation points the fluctuation range of $S(t_k)$ grows significantly, and the clustering of points becomes asymmetric due to biased residence times in each lobe. 
This behavior highlights how the onset of chaos is accompanied by measurable changes in the distributions of $H_1$ and $S$. 
\color{black}
The NNET section samples therefore provide a complementary thermodynamic-geometric characterization, rather than an independent proof of chaos.
\color{black}
\subsection{Chen System}
In this subsection, we analyze the Chen system\cite{key-999}, which exhibits behavior similar to the Lorenz system\cite{key-lorenz}, a representative dynamical system with chaotic behavior. Unlike the Lorenz system, where the phase-space contraction rate is constant and bifurcations produce abrupt changes, the Chen system exhibits a parameter-dependent contraction rate, leading to continuous deformations of the phase-space structure. This manifests in the NNET section analysis as gradual distortions and asymmetries in the $(H_1, S)$ distribution, rather than sharp transitions.

Like the Lorenz system, the Chen system can be described as a toy model consisting of the following set of three differential equations:
\begin{eqnarray}
\dot{x} &=& a(y - x), \\
\dot{y} &=& (c-a)x+cy-xz, \\
\dot{z} &=& xy - b z,
\end{eqnarray}
where we set the parameters as $a = 35$, $b = 3$ with the time evolution computed numerically using the Dormand–Prince method. While the Lorenz system exhibits a constant rate of phase-space volume contraction (uniformly dissipative), the Chen system's contraction rate varies with the parameter $c$, resulting in a continuous change.
\color{black}
For the initial condition $x(0)=y(0)=z(0)=1$, the trajectories at $c=18$ and
$19$ converge to stable nonzero equilibria after damped oscillations.
\color{black}
\begin{figure}[H]
\centering
\includegraphics[width=1.0\textwidth,height=0.45\textwidth]{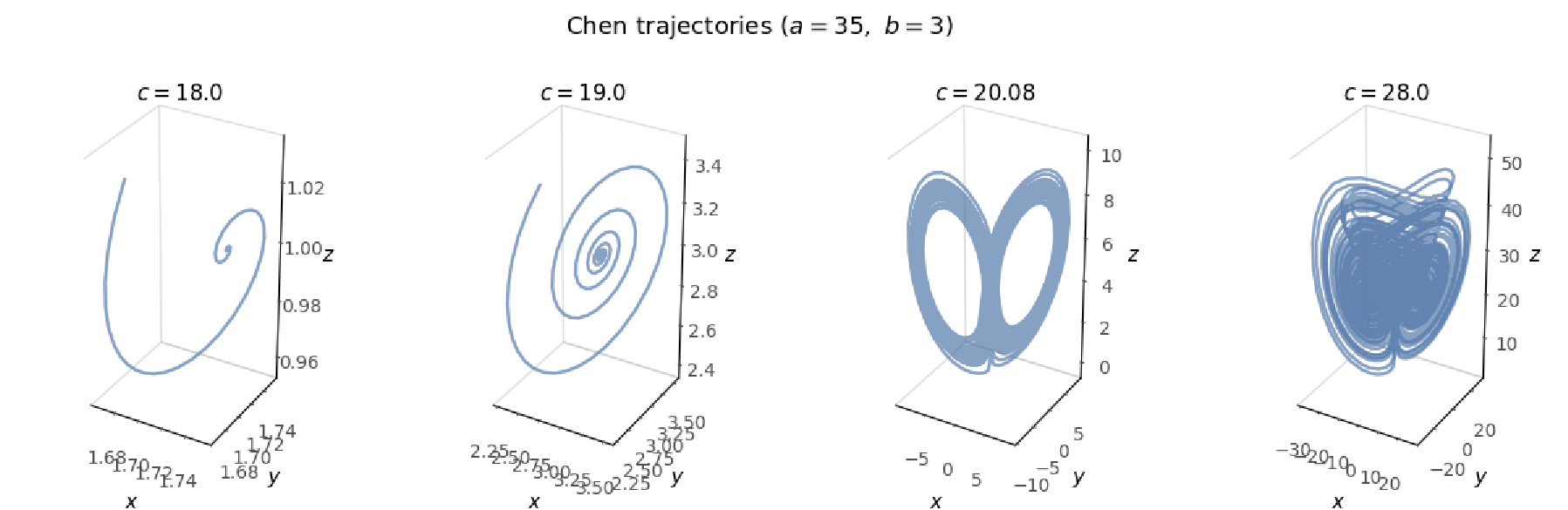}
\caption{Trajectory of the Chen system with initial condition $x(0) = 1.0$, $y(0) = 1.0$, $z(0) = 1.0$, and $dt_{init} = 0.005$. \color{black}The cases $c=18$ and $19$ show damped oscillations toward stable nonzero
equilibria, whereas $c=20.08$ and $28$ exhibit chaotic dynamics for the
stated initial condition.}
\label{aa}
\end{figure}

Within the framework of NNET, we introduce the following expressions for the Hamiltonians $H_1$, $H_2$, and the entropy $S$:
\begin{eqnarray}
H_1 &=& \frac{1}{2}(2a-c)(y^2+z^2), \\
H_2 &=& -z + \frac{1}{2(2a-c)}x^2, \\
S &=& -\frac{1}{2}ax^2 + \frac{1}{2}cy^2 - \frac{1}{2}bz^2 + (c-a) xy
\end{eqnarray}
where $2a-c > 0$.

\begin{figure}[H]
\centering
\includegraphics[width=0.65\textwidth,height=0.45\textwidth]{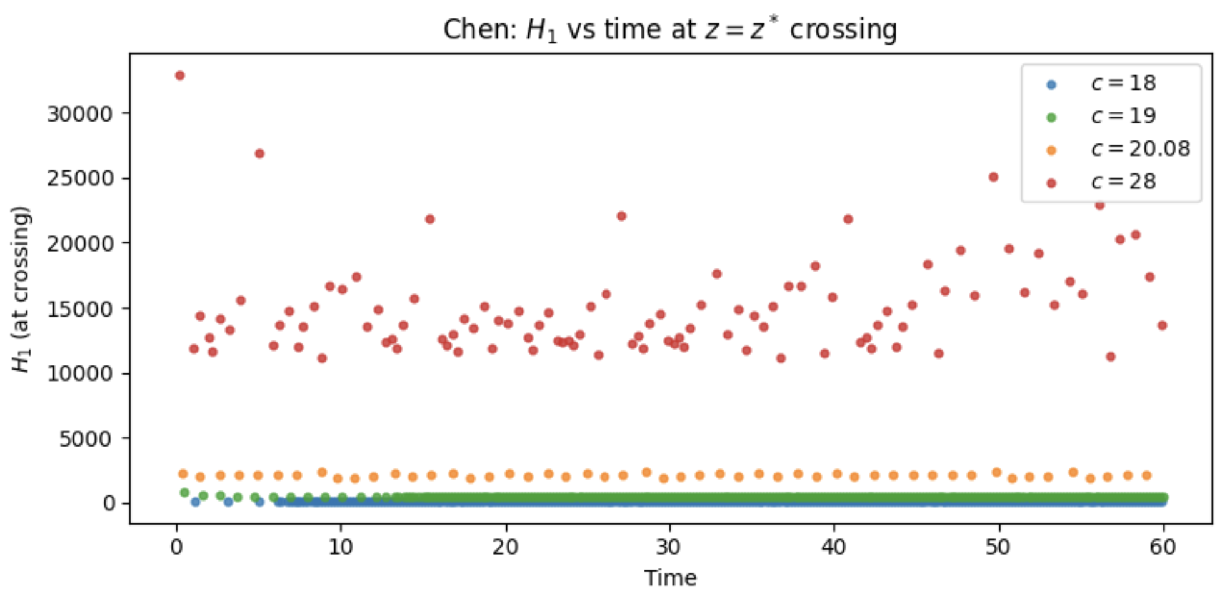}
\caption{Time evolution of the Hamiltonian $H_1$ for the Chen system over the time interval $0 \le t \le 60$ and $dt_{init} = 0.005$ with initial condition $x(0) = 1.0$, $y(0) = 1.0$, and $z(0) = 1.0$.}
\label{bb}
\end{figure}

\begin{figure}[H]
\centering
\includegraphics[width=0.65\textwidth,height=0.45\textwidth]{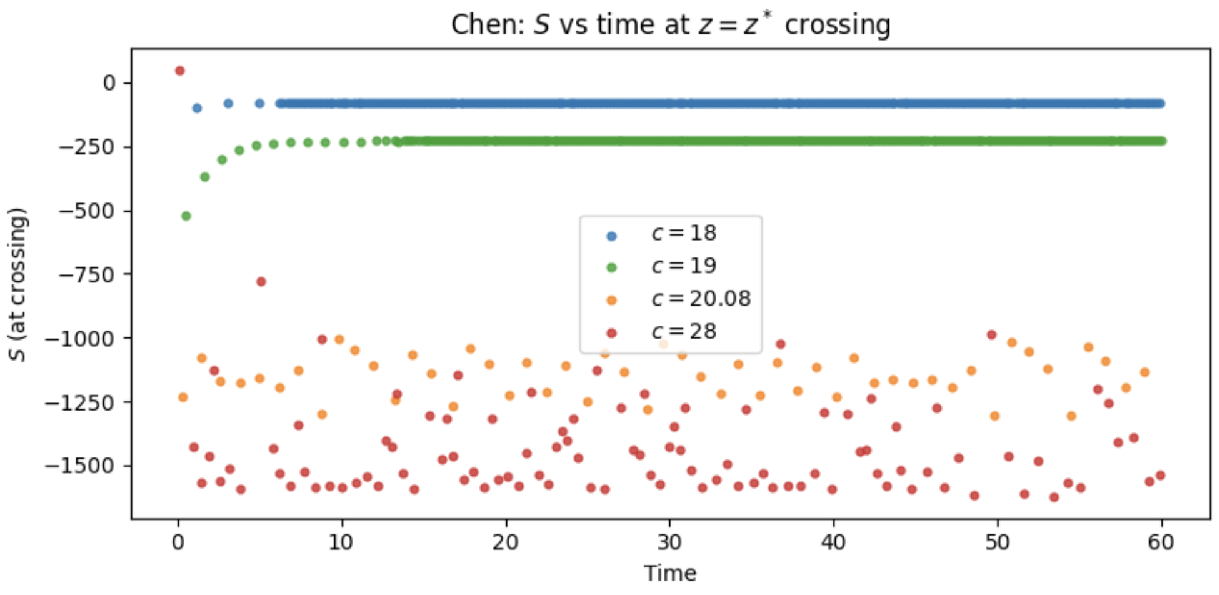}
\caption{Time evolution of the Hamiltonian $S$ for the Chen system over the time interval $0 \le t \le 60$ and $dt_{init} = 0.005$ with initial condition $x(0) = 1.0$, $y(0) = 1.0$, and $z(0) = 1.0$.}
\label{dd}
\end{figure}
Figure \ref{aa} shows the typical attractor of the Chen system in three-dimensional phase space.
As in the Lorenz case, in Figures~\ref{bb}, \ref{dd} we evaluate the section values
$H_1(t_k)$ and $S(t_k)$ on the plane $z=z^\ast:=2c-a$.
As $c$ increases, the distribution changes from a narrow unimodal cluster
to broader multimodal bands, indicating the transition from stable motion
to chaotic dynamics.

It is worth noting that in the Lorenz system, dissipation is constant, so the time evolution of $S$ is highly sensitive to changes in the threshold value of $\rho$, and the $(H_1,S)$ patterns change dramatically at each bifurcation. In contrast, in the Chen system the dissipation rate depends explicitly on $c$, and thus the variations in $(H_1,S)$ evolve more continuously, appearing as asymmetry in orbital residence times and clustering on the attractor.


\color{black}
\section{Summary and Discussion}\label{sec6}

We have shown that Nambu Non-equilibrium Thermodynamics (NNET) gives a practical and unified description of far-from-equilibrium dynamics across four paradigmatic systems: the BZ reaction, the Hindmarsh–Rose neuron, and the Lorenz and Chen chaotic flows.
 In all cases the decomposition into a multi-Hamiltonian (incompressible) flow and a dissipative gradient flow cleanly separates structure-forming dynamics from dissipation.

For a Hamiltonian $H_a$, the time derivative under the metric-compatible NNET dynamics is $\dot H_a = -\{H_1,H_2,H_a\}_{\rm NB} + g^{ij}\partial_i H_a\,\partial_j S$ .  If $H_a$ is one of the Nambu Hamiltonians, the first term vanishes by the antisymmetry of the Nambu bracket. Thus, the Hamiltonians are conserved by the Nambu sector. In the full dynamics, however, the entropy-gradient term can change $H_a$ unless the metric-orthogonality condition $g^{ij}\partial_i H_a\,\partial_j S =0$ is satisfied. Therefore, Hamiltonian energy should be understood as an exact invariant of the non-dissipative Nambu sector and as a diagnostic quantity in the full dissipative system. The drift of $H_a$ measures the coupling between the conservative Nambu structure and dissipation or external reservoirs. The present approach does not aim to replace such diagnostics. Rather, its advantage is that the Hamiltonian quantities are embedded into a thermodynamic decomposition of the full vector field, $\dot{x}^{i} = -\{H_1,H_2,x^{i}\}_{\rm NB} + g^{ij}\partial_j S$.  This representation separates the non-dissipative Nambu contribution from the dissipative entropy-gradient contribution. Consequently, variations of Hamiltonian quantities in the full system can be interpreted as Hamiltonian drift induced by dissipative or reservoir coupling. In chaotic regimes, the joint behavior of Hamiltonian and entropy-like variables, for example through Poincare-section samples $\{H_1(t_k),S(t_k)\}$, provides a state-space diagnostic of the transition from stationary or periodic behavior to chaotic behavior. pectral entropy is a signal-processing measure computed from the normalized power spectrum of a time series. It quantifies how broadly the signal power is distributed over frequency modes and is therefore useful for measuring broadband temporal complexity in chaotic signals\cite{Ino}. By contrast, the entropy-like potential $S$ in NNET is not defined from the Fourier spectrum of a signal. It is a scalar potential on the state space and generates the dissipative part of the vector field through $g^{ij}\partial_j S$. Thus, spectral entropy and the NNET entropy-like potential are complementary. Spectral entropy measures frequency-domain complexity of an observed time series, whereas $S$ describes the state-space structure of dissipation. In chaotic systems, spectral entropy can quantify temporal irregularity, while the Hamiltonian--entropy representation in NNET helps reveal how trajectories move through the conservative--dissipative geometry of the attractor.

\color{black}
For the BZ reaction, the dynamics require multiple Hamiltonians together with an explicit entropic term. Along a cycle, entropic and Hamiltonian contributions almost cancel, concentrating entropy changes into short ``kicks''; this clarifies why Onsager's near-equilibrium theory fails here and how oscillations persist despite dissipation. For the Hindmarsh–Rose model, the zero-set conditions for $\dot H_1,\dot H_2,\dot S$ identify phase-flip points on the limit cycle.
Near these intersections, $H_1$ behaves approximately as a pseudo-conserved quantity, which accounts for the timing of spikes, while the slow dynamics of $H_2=z$ with $\dot H_2 = r{s(x_R - x)+z}$ explain its staircase-like increase.

Although we focused on spatially uniform ODEs, the construction readily extends to spatially extended systems.
This suggests new analyses of entropy--Hamiltonian interplay in pattern formation and other nonlinear media.
We hope that the present results will contribute to a more unified understanding of far-from-equilibrium oscillatory, spiking, and chaotic dynamics\footnote{e.g., Turing\cite{key-tur}, reaction–diffusion\cite{key-rd1,key-rd2}) and other oscillatory or chaotic media (e.g., Bray–Liebhafsky (BL) model\cite{key-1}, lasers\cite{key-other1}, astrophysics\cite{key-other2}, earthquakes\cite{key-other3}, engineering applications\cite{key-other4,key-other5}, and psychology\cite{key-other6}}.


\section*{Acknowledgments}
We thank Toshio Fukumi for helpful discussions on nonlinear response theory. We are indebted to Shiro Komata for reading this paper and
giving useful comments. 


\appendix
\def\thesection{Appendix \Alph{section}}

\section{Analysis of cycles and spikes}\label{sec4}
\begin{figure}[H]
    \begin{tabular}{cc}
      \begin{minipage}[t]{0.45\hsize}
        \centering
        \includegraphics[width=1.0\textwidth,height=1.0\linewidth]{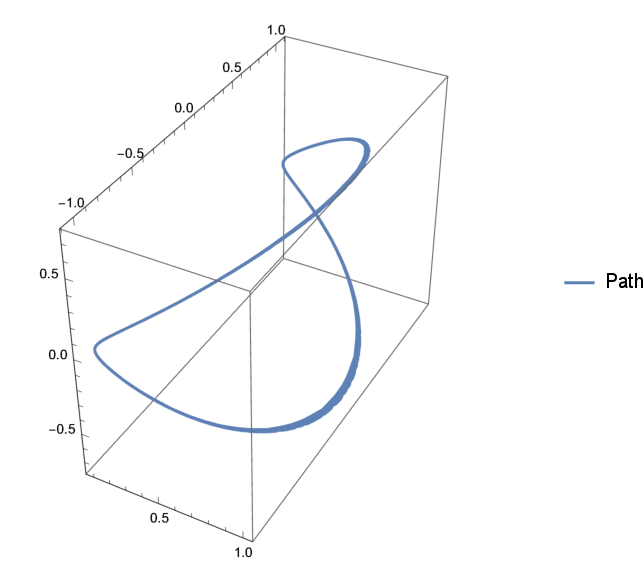}
        \caption{3D diagram of the limit cycle according to the time variation of $x$, $y$, and $z$.}
        \label{fig:spikepath}
      \end{minipage} &
      \begin{minipage}[t]{0.45\hsize}
        \centering
        \includegraphics[width=1.0\textwidth,height=1.0\linewidth]{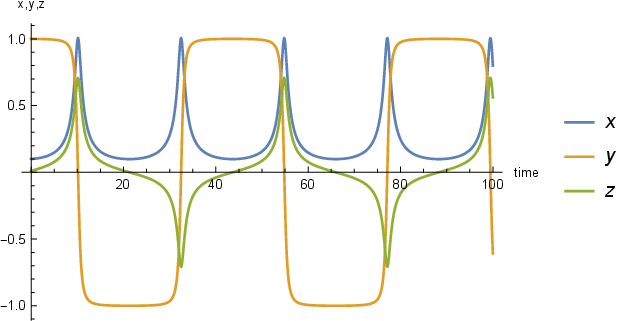}
        \caption{Plot of $x$, $y$ and $z$ as a function of time.}
        \label{fig:spikexyz}
      \end{minipage}
    \end{tabular}
  \end{figure}

Two characteristic behaviors of systems far from equilibrium are cyclic trajectories and the occurrence of spikes. 
When conserved quantities exist, the trajectories are expected to form cycles, as they must continually evolve to satisfy the conservation laws. 
Even in the absence of exact conserved quantities, cycles can persist over long timescales if there exist pseudo-conserved quantities that decrease only asymptotically. 
The triangular reaction\cite{paperI} has already been presented as an example of such a system with a conserved quantity; here, however, we also consider the following simple toy model.
\begin{equation}
H_{1}=a-x^{3}z,
\end{equation}
\begin{equation}
H_{2}=b+z,
\end{equation}
\begin{equation}
S=x^{2}yz.
\end{equation}
The equations of motion are as follows
\begin{equation}
\dot{x}=2xyz,
\end{equation}
\begin{equation}
\dot{y}=-2x^{2}z,
\end{equation}
\begin{equation}
\dot{z}=x^{2}y.
\end{equation}
This preserves 
\begin{equation}
O=\frac{1}{2}(x^{2}+y^{2})
\end{equation}
in time evolution.
The path draws a limit cycle as in Figures \ref{fig:spikepath} and \ref{fig:spikexyz}, with a sharp spike for the $x$ and $z$ components resulting from its nonlinearity.

Another toy model is 
\begin{equation}
    H_1 = H_1(x,y,z),
\end{equation}
\begin{equation}
    H_2 = (x^2 - y^2) f(z),
\end{equation}
\begin{equation}
    S = xy,
\end{equation}
where $H_1(x,y,z)$ and $f(z)$ are arbitrary functions. The equations of motion are
\begin{equation}
\dot{x}=y-2yf(z)\frac{\partial H_{1}}{\partial z}+(y^{2}-x^{2})\frac{df(z)}{dz}\frac{\partial H_{1}}{\partial y},
\end{equation}
\begin{equation}
\dot{y}=x-2xf(z)\frac{\partial H_{1}}{\partial z}-(y^{2}-x^{2})\frac{df(z)}{dz}\frac{\partial H_{1}}{\partial x},
\end{equation}
\begin{equation}
\dot{z}=2xf(z)\frac{\partial H_{1}}{\partial y}+2yf(z)\frac{\partial H_{1}}{\partial x}.
\end{equation}

In this model, $H_2$ itself is the conserved quantity of time evolution.

As a concrete example, we take
\begin{equation}
    H_1 = \frac{1}{2}(x^2 + y^2 + z^2),
\end{equation}
\begin{equation}
    f(z) = z^2.
\end{equation}

The equations of motion are
\begin{equation}
\dot{x}=y-2x^{2}yz+2y^{3}z-2yz^{3},
\label{eq:spikeToyModel}
\end{equation}
\begin{equation}
\dot{y}=x+2x^{3}z-2xy^{2}z-2xz^{3},
\label{eq:spikeToyModel2}
\end{equation}
\begin{equation}
\dot{z}=4xyz^{2}.
\label{eq:spikeToyModel3}
\end{equation}

\begin{figure}[H]
    \begin{tabular}{cc}
      \begin{minipage}[t]{0.45\hsize}
        \centering
        \includegraphics[width=1.0\textwidth,height=1.0\linewidth]{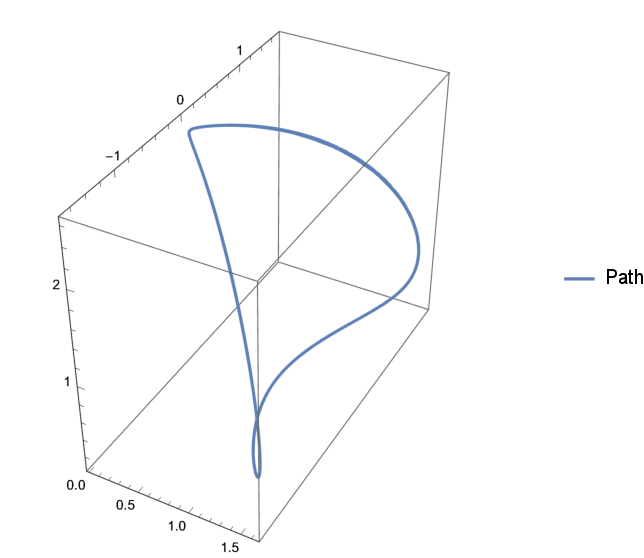}
        \caption{3D diagram of the limit cycle according to the time variation of $x$, $y$, and $z$.}
        \label{fig:h1spikepath}
      \end{minipage} &
      \begin{minipage}[t]{0.45\hsize}
        \centering
        \includegraphics[width=1.0\textwidth,height=1.0\linewidth]{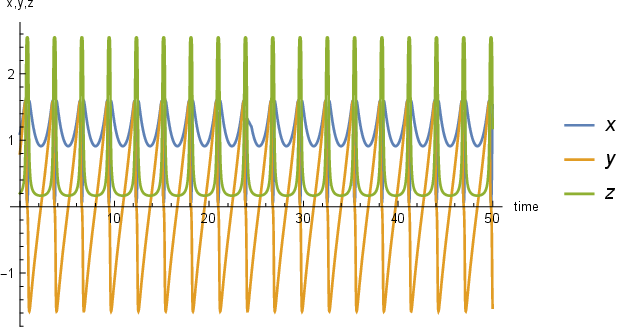}
        \caption{Plot of $x$, $y$, and $z$ as a function of time.}
        \label{fig:h1spikexyz}
      \end{minipage}
    \end{tabular}
  \end{figure}

The path draws a limit cycle as in Figures \ref{fig:h1spikepath} and \ref{fig:h1spikexyz}, with a sharp spike for the $y$ and $z$ components resulting from its nonlinearity.

\section{\color{black}A conceptual linear toy model for oscillatory NNET dynamics\color{black}}\label{secondappendix}
\color{black}
At the end of Subsection \ref{sec3.1}, a simplified truncation of the NNET representation of the BZ reaction was introduced. The purpose of the present appendix is not to derive the periodic orbit of the full Oregonator quantitatively, but to present a conceptual linear toy model illustrating how oscillator-type motion may arise from the interplay between the Nambu and gradient sectors. We therefore consider the following simplified Hamiltonians and entropy-like potential:
\color{black}
\begin{align}
    H_1 &= a Y Z, \\
    H_2 &= Z, \\
    S &= \frac{1}{2} b X^2 + c X Z.
\end{align}

The evolution of the system follows the Nambu non-equilibrium thermodynamic equation:
\begin{align}
    \dot{X}^i &= - \{ H_1, H_2, X^i \} + \frac{\partial S}{\partial X^i}.
\end{align}

For the variables \(X\), \(Y\), and \(Z\), this gives the following equations of motion:
\begin{align}
    \dot{X} &= - a Z + b X + c Z = (c - a) Z + b X, \\
    \dot{Y} &= 0, \\
    \dot{Z} &= c X. \label{H.O.}
\end{align}

We can express this system in matrix form as:
\begin{align}
    \begin{pmatrix}
        \dot{X} \\
        \dot{Z}
    \end{pmatrix}
    &= 
    \begin{pmatrix}
        b & c - a \\
        c & 0
    \end{pmatrix}
    \begin{pmatrix}
        X \\
        Z
    \end{pmatrix}.
\end{align}

Differentiating once more gives:
\begin{align}
    \begin{pmatrix}
        \ddot{X} \\
        \ddot{Z}
    \end{pmatrix}
    = 
    \begin{pmatrix}
        b & c - a \\
        c & 0
    \end{pmatrix}^2
    \begin{pmatrix}
        X \\
        Z
    \end{pmatrix}
    =
    \begin{pmatrix}
        b^2 + c(c - a) & b(c - a) \\
        bc & c(c - a)
    \end{pmatrix}
    \begin{pmatrix}
        X \\
        Z
    \end{pmatrix}.
\end{align}

The eigenvalue equation of this matrix,
\begin{align}
\lambda^2-\bigl(b^2+2c(c-a)\bigr)\lambda+c^2(c-a)^2=0. 
\end{align}
Since $\lambda_+\lambda_-=\det(M^2)=c^2(c-a)^2\ge0$, both eigenvalues have the same sign. 
In particular, when $b^2+2c(c-a)<0$ (i.e. $c(c-a)<-b^2/2$), both eigenvalues are negative, 
and the corresponding eigenmode is oscillatory (elliptic). In other words, \color{black}this simplified construction should be interpreted as a conceptual illustration of the possibility that an oscillator-type sector and an approximate quadratic invariant may emerge after a suitable reduction or reorganization of the NNET variables. It does not establish an exact hidden conservation law for the full Oregonator system. Whether such a reduced invariant accurately describes the BZ limit cycle depends on the validity of the truncation and must be tested independently. In particular, the local onset of oscillation should be analyzed through the Jacobian spectrum near a stationary point, while the stability of the resulting periodic orbit requires a Floquet-multiplier analysis.

\color{black}
%
%
%

\section{Turning points of spikes}\label{firstappendix}
In this section, we discuss the turning points of spikes in the toy model in Section \ref{sec2}.

From Eq. (\ref{eq:spikeToyModel})--(\ref{eq:spikeToyModel3}), we obtain the following equations,
\begin{align}
    dt &= \frac{dx}{y-2x^{2}yz+2y^{3}z-2yz^{3}} \\
       & = \frac{dy}{x+2x^{3}z-2xy^{2}z-2xz^{3}} \\
        & = \frac{dz}{4xyz^{2}}.
\end{align}
Now consider a cusp that changes abruptly only in the $x$ direction.
Since the denominator is quadratic for $x$, the equation can be rearranged as follows.
\begin{equation}
    dt = \frac{dx}{(x-\alpha_+)(x-\alpha_-)} = dx\left(\frac{A_+}{x-\alpha_+}+\frac{A_-}{x-\alpha_-}\right),
\end{equation}
\begin{equation}
    \alpha_\pm \equiv \mp \sqrt{y^2-z^2+\frac{1}{2 z}},
\end{equation}
\begin{equation}
    A_\pm = 2\alpha_\pm.
\end{equation}

Since $dt$ is singular in terms of the cusp arising and 
only the contribution of the $A_+$ term works around $x = \alpha_+$,
 we can write the following equation.

\begin{equation}
    \frac{x(t)-\alpha_+}{x(t_0) - \alpha_+} = e^{(t-t_0)/A_+} = (t-t_0)/A_+ + O((t-t_0)^2/A_+^2)
\end{equation}

Therefore, the gap before and after the turning point can be estimated from the following equation:
\begin{align}
    \dot{x}(t_* + \delta t) - \dot{x}(t_* - \delta t) &= 1/A_+ \left(x(t_* + \delta t) + x(t_* - \delta t)\right) \\
     & = \frac{1}{2\sqrt{y(t_*)^2-z(t_*)^2+\frac{1}{2 z(t_*)}}}\left(x(t_* + \delta t) + x(t_* - \delta t)\right),
\end{align}
where $t_*$ is time at the turning point.




\begin{thebibliography}{999}
\bibitem{degroot1962}S. R. de Groot and P. Mazur, ``Non-Equilibrium Thermodynamics,'' North-Holland Publishing Co. Amsterdam, Interscience Publishers (1964).

\bibitem{Prigogine}P. Glansdorff and I. Prigogine, ``On a general evolution criterion in macroscopic physics,'' Physica, 30(2):351 (1964).

\bibitem{Prigogine2}G. Nicolis and I. Prigogine, ``Self-Organization in Nonequilibrium Systems: From Dissipative Structures to Order through Fluctuations,'' Wiley  (1977).

\bibitem{key-203}H. Haken, ``Cooperative phenomena in systems far from thermal equilibrium and in nonphysical systems,'' Rev. Mod. Phys. 47, 67 (1975).

\bibitem{key-50}M. Grmela and H. C. {\"O}ttinger, ``Dynamics and thermodynamics of complex fluids. I. Development of a general formalism,'' Physical Review E 56.6 6620 (1997).
\bibitem{key-200}H. C. {\"O}ttinger, and M. Grmela, ``Dynamics and thermodynamics of complex fluids. II. Illustrations of a general formalism,'' Physical Review E 56.6 6633 (1997).

\bibitem{key-201}H. C. {\"O}ttinger, ``Beyond Equilibrium Thermodynamics,'' Wiley, Hoboken (2004).

\bibitem{key-202}M. Grmela, ``GENERIC guide to the multiscale dynamics and thermodynamics,'' Journal of Physics Communications 2.3 032001 (2018).


\bibitem{key-300}A. T. Winfree, ``The Geometry of Biological Time,'' Springer (1980).

\bibitem{key-301}Y. Kuramoto, ``Chemical Oscillations, Waves, and Turbulence,'' Springer (1984).

\bibitem{key-302}E. Brown, J. Moehlis, and P. Holmes, ``On the Phase Reduction and Response Dynamics of Neural Oscillator Populations,'' Neural Computation 16 (4) 673–715 (2004).

\bibitem{key-111}H. Nakao, ``Phase reduction approach to synchronization of nonlinear oscillators,'' Contemporary Physics 57, 188-214  (2016), arXiv:1704.03293 [nlin].
\bibitem{key-222}S. Shirasaka, T. Kurebayashi, and H. Nakao, ``Phase-amplitude reduction of transient dynamics far from attractors for limit-cycling systems,'' Chaos, 27 (2) (2017), arXiv:1701.05428 [nlin].

\bibitem{key-2222}A. Mauroy,  I. Mezi{\'c}, and Y. Susuki, ``The Koopman Operator in Systems and Control,'' Springer (2020).



\bibitem{key-2.2}E. Ott and T. M. Antonsen, ``Low Dimensional Behavior of Large Systems of Globally Coupled Oscillators,'' Chaos, 18 037113 (2008), arXiv:0806.0004 [nlin].
\bibitem{key-tur}A. M. Turing, ``The chemical basis of morphogenesis,'' Philos. Trans. R. Soc. Lond. B Biol. Sci. 237, 37–72 (1952).

\bibitem{key-0.1}M. C. Cross and P. C. Hohenberg, ``Pattern formation outside of equilibrium,'' Rev. Mod. Phys. 65, 851 (1993).

\bibitem{key-0.2}M. Cross and H. Greenside, ``Pattern Formation and Dynamics in Nonequilibrium Systems,'' Cambridge Univ. Press (2009).

\bibitem{key-0.3}R. J. Field and E. K{\"o}r{\"o}s and R. M. Noyes, ``Oscillations in chemical systems. II. Thorough analysis of BZ-type systems,'' J. Am. Chem. Soc. 94 25 (1972).

\bibitem{key-0.4}R. J. Field and M. Burger, ``Oscillations and Traveling Waves in Chemical Systems,'' Wiley (1985).

\bibitem{key-0.5}I. R. Epstein and J. A. Pojman, ``An Introduction to Nonlinear Chemical Dynamics,'' Oxford Univ. Press (1998).


\bibitem{key-one}A. L. Hodgkin and A. F. Huxley, ``A quantitative description of membrane current and its application to conduction and excitation in nerve,'' J. Physiol. 117 (4) 500–544 (1952).

\bibitem{key-two}J. Rinzel, ``A formal classification of bursting mechanisms in excitable systems,'' Springer (1987).

\bibitem{key-three}E. M. Izhikevich, ``Dynamical Systems in Neuroscience: The Geometry of Excitability and Bursting,'' MIT Press (2007).
\bibitem{paperI}
S. Katagiri, Y. Matsuoka, and A. Sugamoto, ``Nambu Non-equilibrium Thermodynamics: Axiomatic Formulation and Foundation,'' Chaos 36 (7) 073133 (2025), arXiv:2508.00207 [cond-mat].
\bibitem{paper0}
S. Katagiri, Y. Matsuoka, and A. Sugamoto, ``Fluctuating Non-linear Non-equilibrium System in Terms of Nambu Thermodynamics," (2022), arXiv:2209.08469 [cond-mat].


\bibitem{key-9}Y. Nambu, ``Generalized Hamiltonian dynamics,'' Physical Review D 7 2405-2412 (1973).

\bibitem{paperII}
S. Katagiri, Y. Matsuoka, and A. Sugamoto, ``Reduction of Complex Dynamics in Far-from-equilibrium Systems: Nambu Non-equilibrium Thermodynamics,'' 	J. Math. Phys. 67, 042704 (2026), arXiv:2508.19455 [cond-mat].
\bibitem{key-b}B. P. Belousov, ``A periodically acting reaction and its mechanism,'' Collection of Abstracts on Radiation Medicine, 147, 145 (1959). [In Russian.]

\bibitem{key-z}A. M. Zhabotinsky, ``Periodic process of oxidation of malonic acid in solution,'' Biophysics (Biofizika), 9, 306–311 (1964). [In Russian.]


\bibitem{key-51}J. L. Hindmarsh and R. M. Rose. ``A model of neuronal bursting using three coupled first order differential equations,'' Proceedings of the Royal society of London. Series B. Biological sciences 221.1222 87-102 (1984).

\bibitem{key-lorenz}E. N. Lorenz, ``Deterministic nonperiodic flow,'' Journal of the Atmospheric Sciences 20 130-141 (1963).

\bibitem{key-999}G. Chen and T. Ueta, ``Yet another chaotic attractor,'' Int. J. Bifurc. Chaos 9 1465–1466 (1999).

\bibitem{key-3}L. Onsager, ``Reciprocal relations
in irreversible processes. I.,''{} Physical Review 37.4 405
(1931).

\bibitem{metri}P. J. Morrison, ``Bracket formulation for irreversible classical fields,'' Physics Letters A 100(8), 423–427, (1984).
\color{black}
\bibitem{ad1}Pavelka, Klika, Grmela, ``Multiscale Thermo-Dynamics, de Gruyter,'' textbook (2018).
\bibitem{ad2}R. Kraaij, A. Lazarescu, C. Maes, and M. Peletier, ``Deriving GENERIC from a generalized fluctuation symmetry. Journal of Statistical Physics,'' Journal of Statistical Physics 170, 492-508 (2018), arXiv:1706:10115[cond-mat].

\bibitem{ad3}S. Katagiri, ``Nambu Nonequilibrium Thermodynamics and the Lyapunov Structure of Open Systems,'' arXiv:2606.05231[cond-mat] (2026).

\color{black}
\bibitem{key-52}B. Bao, et al. ``Three-dimensional memristive Hindmarsh-Rose neuron model with hidden coexisting asymmetric behaviors,'' Complexity 2018 (2018).

\bibitem{key-t}W. Tucker. ``The Lorenz attractor exists,'' C. R. Acad. Sci. Paris S\'er. I Math., 328(12):1197–1202, (1999).

\bibitem{key-j}S. Jafari, J. Sprott, and F. Nazarimehr, ``Recent new examples of hidden attractors,'' The European Physical Journal Special Topics 224 8 1469–1476 (2015).
\bibitem{AF}M. Axenides and E. Floratos, ``Strange attractors in dissipative Nambu mechanics: classical and quantum aspects,'' JHEP. 1004, 36 (2010), arXiv:0910.3881 [nlin].

\bibitem{Ino}T. Inouye, K. Shinosaki, H. Sakamoto, S. Toi, S. Ukai, A. Iyama, Y. Katsuda, and M. Hirano, ``Quantification of EEG irregularity by use of the entropy of the power spectrum,'' Electroencephalography and Clinical Neurophysiology 79(3), 204–210, (1991).

\bibitem{key-rd1}M.C. Milinkovitch, E. Jahanbakhsh, and S. Zakany, ``The unreasonable effectiveness of reaction diffusion in vertebrate skin color
patterning,'' Annu. Rev. Cell Dev. Biol. 39, 145–174 (2023). 

\bibitem{key-rd2}V. Volpert, and S. Petrovskii, ``Reaction-diffusion waves in biology: new trends, recent developments,'' Physics of Life Reviews 6 267–310 (2009).

\bibitem{key-1}R. Vilcu, D. D. Tiberiu, and B. Daniela, ``The Study of Bray-Liebhafsky Reaction Over a Wide Range of Temperatures. II. Modelling,'' Discrete Dynamics in Nature and Society 4 55-62 (2000).

\bibitem{key-other1}M. J. Zhang and Y. C. Wang, ``Review on chaotic lasers and measurement applications,'' J. Lightwave Technol. 39 12, 3711–3723 (2021).


\bibitem{key-other2}F. K. Peng, F. Y. Wang, X. W. Shu, and S. J. Hou, ``Self-organized criticality in solar GeV flares,'' MNRAS 518 3959 (2023).

\bibitem{key-other3}T. Chelidze, G. Melikadze, T. Kiria, T. Jimsheladze, and G. Kobzev, ``Statistical and Non-linear Dynamics Methods of Earthquake Forecast: Application in the Caucasus,'' Front. Earth Sci. 8  194 (2020).

\bibitem{key-other4}S. Gao, L. Chang, I. Romic, Z. Wang, M. Jusup, and P. Holme, ``Optimal control of networked reaction–diffusion systems,'' Journal of the
Royal Society Interface 19  20210739 (2022).

\bibitem{key-other5}J. Gorecki, et al., ``Chemical computing with reaction–diffusion processes,'' Phil. Trans. R. Soc. A 373 20140219 (2015).

\bibitem{key-other6}E. Tognoli, et al., ``Coordination dynamics: A foundation for understanding social behavior,'' Frontiers in Human Neuroscience 14 317 (2020).
\end{thebibliography}
\end{document}